\documentclass[fleqn,usenatbib]{mnras}

\usepackage{newtxtext,newtxmath}

\usepackage[T1]{fontenc}
\usepackage{ae,aecompl}

\usepackage{graphicx}	
\usepackage{amsmath}	
\usepackage{amssymb}	
\graphicspath{{./plots/}}

\newcommand{\msun}{M$_{\odot}$}
\newcommand{\msunyr}{M$_{\odot}$~yr$^{-1}$}
\newcommand{\rsun}{R$_{\odot}$}
\newcommand{\lsun}{L$_{\odot}$}
\newcommand{\kms}{km~s$^{-1}$}
\newcommand{\ergs}{erg s$^{-1}$}
\newcommand{\Ha}{H$\alpha$}

\newcommand{\HeI}{He~{\sc i}}
\newcommand{\HeII}{He~{\sc ii}}
\newcommand{\OI}{O~{\sc i}}

\newcommand{\NII}{N~{\sc ii}}

\newcommand{\Oneb}{[O~{\sc i}]}

\newcommand{\NaI}{Na~{\sc i}}

\newcommand{\MgI}{Mg~{\sc i}}

\newcommand{\SiII}{Si~{\sc ii}}

\newcommand{\CaII}{Ca~{\sc ii}}

\newcommand{\FeII}{Fe~{\sc ii}}

\newcommand{\Cofs}{$^{56}$Co}
\newcommand{\Nifs}{$^{56}$Ni}
\newcommand{\mej}{$M_\mathrm{ej}$}

\newcommand{\vsc}{$v_\mathrm{sc}$}

\newcommand{\lam}{$\lambda$}

\newcommand{\Eh}{$E\left(B-V\right)_\mathrm{host}$}
\newcommand{\Emw}{$E\left(B-V\right)_\mathrm{MW}$}
\newcommand{\Etot}{$E\left(B-V\right)_\mathrm{tot}$}

\newcommand{\mni}{$M_\mathrm{Ni}$}

\newcommand{\texp}{$t_\mathrm{exp}$}

\newcommand{\nel}{$n_e$}
\newcommand{\Mc}{$M_\mathrm{c}$}
\newcommand{\Me}{$M_\mathrm{e}$}
\newcommand{\ve}{$v_\mathrm{e}$}

\newcommand{\rph}{$R_\mathrm{ph}$}
\newcommand{\sn}{SN\,}

\title[SN 2018gjx]{SN 2018gjx reveals that some SNe Ibn are SNe IIb exploding in dense circumstellar material}

\author[]{S. J. Prentice,$^{1}$\thanks{E-mail: sipren.astro@gmail.com}, K. Maguire$^{1}$, I. Boian$^{1}$, J. Groh$^{1}$, J.~Anderson$^{2}$, 
C.~Barbarino$^{3}$, \newauthor K.~A.~Bostroem$^{4}$, J. Burke$^{5,6}$, P.~Clark$^{1,7}$, Y.~Dong$^{4}$, M. Fraser$^{8}$,  L. Galbany$^{9}$,\newauthor   M.~Gromadzki$^{10}$, C.~P.~Guti\'errez$^{11}$, D. A. Howell$^{5,6}$, D.~ Hiramatsu$^{5,6}$, C. Inserra$^{12}$,  \newauthor P.~A.~James$^{13}$, E.~Kankare$^{14}$,  H. Kuncarayakti$^{14,15}$, P.~A.~Mazzali$^{13,16}$, C. McCully$^{5,6}$, \newauthor T.~E.~M\"uller-Bravo$^{11}$, 
 M.~Nichol$^{17,18}$,  C. Pellegrino$^{5,6}$,  S.~J.~Smartt$^{7}$, J. Sollerman$^{3}$, \newauthor L.~Tartaglia$^{19,3}$, S. Valenti$^{4}$,  and D.~R.~Young$^{7}$.
\\
$^{1}$School of Physics, Trinity College Dublin, The University of Dublin, Dublin 2, Ireland\\
$^{2}$European Southern Observatory, Alonso de C\'ordova 3107, Casilla 19, Santiago, Chile\\
$^{3}$The Oskar Klein Centre, Department of Astronomy, Stockholm University, AlbaNova, SE-106 91 Stockholm , Sweden\\
$^{4}$Department of Physics, University of California, 1 Shields Avenue, Davis, CA 95616-5270, USA\\
$^{5}$Las Cumbres Observatory, 6740 Cortona Drive, Suite 102, Goleta, CA 93117-5575, USA\\
$^{6}$Department of Physics, University of California, Santa Barbara, CA 93106-9530, USA\\
$^{7}$Astrophysics Research Centre, School of Mathematics and Physics, Queen's University Belfast, BT7 1NN, UK\\
$^{8}$School of Physics, O'Brien Centre for Science North, University College Dublin, Belfield, Dublin 4, Ireland\\
$^{9}$Departamento de F\'isica Te\'orica y del Cosmos, Universidad de Granada, E-18071 Granada, Spain\\
$^{10}$Astronomical Observatory, University of Warsaw, Al. Ujazdowskie 4, 00-478 Warszawa, Poland\\
$^{11}$Department of Physics and Astronomy, University of Southampton, Southampton, SO17 1BJ, UK\\
$^{12}$School of Physics \& Astronomy, Cardiff University, Queens Buildings, The Parade, Cardiff, CF24 3AA, UK\\
$^{13}$Astrophysics Research Institute, Liverpool John Moores University, IC2, Liverpool Science Park, 146 Brownlow Hill, \\  Liverpool L3 5RF, UK\\
$^{14}$Tuorla Observatory, Department of Physics and Astronomy, FI-20014 University of Turku, Finland\\
$^{15}$Finnish Centre for Astronomy with ESO (FINCA), FI-20014 University of Turku, Finland\\
$^{16}$Max-Planck-Institut f{\"u}r Astrophysik, Karl-Schwarzschild-Str. 1, D-85748 Garching, Germany\\
$^{17}$Birmingham Institute for Gravitational Wave Astronomy and School of Physics and Astronomy, University of Birmingham, Birmingham B15 2TT, UK \\
$^{18}$Institute for Astronomy, University of Edinburgh, Royal Observatory, Blackford Hill, EH9 3HJ, UK \\
$^{19}$INAF - Osservatorio Astronomico di Padova, Vicolo dell'Osservatorio 5, 35122 Padova, Italy\\
}

\date{Accepted XXX. Received YYY; in original form ZZZ}

\pubyear{2020}

\begin{document}
\label{firstpage}
\pagerange{\pageref{firstpage}--\pageref{lastpage}}
\maketitle

\begin{abstract}
We present the data and analysis of \sn2018gjx, an unusual low-luminosity transient with three distinct spectroscopic phases. Phase I shows a hot blue spectrum with signatures of ionised circumstellar material (CSM), Phase II has the appearance of broad SN features, consistent with those seen in a Type IIb supernova at maximum light, and Phase III is that of a supernova interacting with helium-rich CSM, similar to a Type Ibn supernova.
This event provides an apparently rare opportunity to view the inner workings of an interacting supernova.
The observed properties can be explained by the explosion of a star in an aspherical CSM. 
The initial light is emitted from an extended CSM ($\sim 4000$ \rsun), which ionises the exterior unshocked material. Some days after, the SN photosphere envelops this region, leading to the appearance of a SN IIb. 
Over time, the photosphere recedes in velocity space, revealing interaction between the supernova ejecta and the CSM that partially obscures the supernova nebular phase.
Modelling of the initial spectrum reveals a surface composition consistent with compact H-deficient Wolf-Rayet and LBV stars.
Such configurations may not be unusual, with SNe IIb being known to have signs of interaction so at least some SNe IIb and SNe Ibn may be the same phenomena viewed from different angles or, possibly with differing CSM configurations.
\end{abstract}




\section{Introduction}
The final stages of evolution of a massive star should see significant mass loss through winds, mass ejections, or stripping by a companion.
When these massive stars eventually explode as supernovae (SNe), they should do so in environments that are not pristine \citep[see][]{Langer2012,Georgy2012,Groh2014,Smith2014}
so the ejecta should interact with some amount of circumstellar material (CSM). For strongly interacting events, the spectra are dominated by narrow emission lines on a blue continuum, with H-interacting events known as Type IIn and H-poor/He-rich interacting events known as Type Ibn \cite[e.g.][]{2014ARA&A..52..487S, Pastorello2008b}. 
The emission lines can be seen to have two components; an intermediate component measuring $\sim$ a few thousand \kms, formed by shocked gas near the CDS, and a narrow component formed from recombination of ionised unshocked material further out \citep[for an overview, see][]{Smith2017}.
In these SNe, a forward and a reverse shock form, with a cool dense shell at the surface of interaction \citep[CDS, for an overview, see][]{Smith2017}. The high density of the CDS leads to reprocessing of photons emitted from within its boundaries, which, in addition to the high luminosities that can be reached, serves to mask the underlying SN event.
Often, the only signatures of the SN itself in H-poor interacting events are broad \CaII\ near-infrared (NIR) and \OI\ \lam7774 emission \citep{Smith2009,Pastorello2015c}. Consequently, the actual explosions that give rise to interacting SNe are often hidden from view.

The progenitors of Type IIn are expected to be stars which have recently undergone mass loss from a H-rich envelope, whereas it is thought that the progenitors of Type Ibn are H-poor Wolf-Rayet stars \citep{Pastorello2007,Pastorello2008b}, or LBV-like stars transitioning to a Wolf-Rayet phase if weak H lines are seen \citep[e.g., SN 2011hw;][]{Smith2012,Sun2020}. 
Binarity may play some role in the in the difference between SNe IIn and Ibn progenitors, and there is good evidence for a binary companion to the progenitor of Type Ibn SN 2006jc \citep{Sun2020}. 
 
Additionally, CSM/ejecta interaction has also been suggested to explain the high luminosities of superluminous-SNe II \citep{2019ARA&A..57..305G}, as well as unusual events such as luminous blue transients \citep[e.g.,][]{Prentice2018b,2019ApJ...872...18M,2020ApJ...895...49H}.
Aside from events with an apparent core-collapse origin there is also a subset of SNe Ia interacting with dense CSM. These Ia-CSM have early time spectra that are similar to that of SNe IIn \citep[e.g.,][]{Inserra2016}.

Type II SNe display strong broad H features in their spectra from a large H envelope, while stripped-envelope SNe (SE-SNe), also known as SNe Ibc, are the explosions of stars that have undergone some degree of stripping of their outer H- and He- envelopes during their pre-explosion lives and are H deficient (Type Ib) or H and He deficient (Type Ic) \citep[][]{Filippenko1997}. 
Apart from the strongly interacting Type IIn and Type Ibn events, most core-collapse SNe do not display obvious photometric or spectroscopic signatures of interaction at optical wavelengths.
If they do, it is often at very early times, observed through a flux excess in their light curves caused by the ejected material running into nearby CSM \citep[e.g.,][]{Morozova2017,Hoss2018}, delayed shock-breakout due to dense very close-by CSM \citep{Foerster2018}, or from signatures in their early spectra such as the presence of \HeII\ and H recombination lines  \citep[``flash-ionisation'';][]{Benetti1994,Galyam2014,Gang2020}. Less directly, the presence of CSM material can be inferred from precursor mass loss events (sometimes known as ``SN impostors''), which result in a burst of increased luminosity in the years prior to explosion and for which the terminal explosion of the star may also be seen \citep{Pastorello2007, Pastorello2013,Mauerhan2013,Margutti2014,Ofek2016,Pastorello2019}. 
The vast majority of SE-SNe, however, show no conspicuous signs of strong interaction in the optical in the first few hundred days, although there are an increasing number of exceptions \citep[e.g.,][]{Mili2015,Margutti2017,Kuncarayakti2017,Mauerhan2018,Chen2018}

Type IIb are SNe that transition from displaying H-rich spectra (`Type II-like') to displaying He-rich spectra (`Type Ib-like') within a few weeks of explosion \citep[e.g.][]{Matheson2000, Arcavi2011,2012ApJ...757...31B}. This suggests that the progenitor star has lost most, but not all, of its H envelope by the time of explosion and based on this mass loss may display signatures of interaction with CSM. Signatures of interaction in the early spectra of Type IIb SNe have been seen, through the presence of narrow \HeII\ and H recombination lines on a hot blue continuum, in a number of Type IIb SNe, when early enough spectra have been obtained and are often referred to as `flash-ionisation' signatures \citep[e.g.][]{Benetti1994,Galyam2014}. These narrow emission features are thought to be due to UV photons ionising unshocked CSM or through interaction of the SN ejecta with nearby CSM \citep{Leonard2000,Chugai2001,Galyam2014,Boian2020}. 

Some SNe IIb display double peaked light curves, where the first peak is assumed to arise from shock cooling of the stellar surface or from nearby dense CSM, with the second peak caused by radioactive \Nifs\ decay \citep{Woosley1994,2012ApJ...757...31B,2015MNRAS.454...95M,Piro2015}.  
Interaction with CSM can also be seen at late times in Type IIb SNe through optical nebular emission lines \citep{Matheson2000b,Mili2012,Maeda2015}, or through X-ray or radio detections \citep{Fransson1996,Margutti2017}. The appearance of CSM interaction in Type IIb SNe generally requires detailed observations very soon after explosion to catch transient narrow emission lines, at late-times, or at X-ray/radio wavelengths.

In this paper we present the observations and analysis of \sn2018gjx. It is an unusual low-luminosity transient that showed narrow H- and He-ionisation signatures produced in nearby CSM in its early spectra, which then evolved through a SN IIb phase with prominent H and He features, before again showing signatures of interaction at late times. At these late times the interaction is consistent with He CSM interaction as in Type Ibn. It represents the first time that a SN has been seen to display normal SE-SNe spectra (IIb-like) and later evolve into a Type Ibn SN. This demonstrates a connection between at least some Type IIb SNe, Type Ibn SNe, as well as highlighting the role that CSM plays in the observed properties of some events and how this links to significant mass-loss from the progenitor system.
In Section~\ref{sec:data} we present the data collection and reduction processes, as well as properties of the host galaxy.
The light curves are shown in Section~\ref{sec:lc}, along with the colour and temperature evolution of the object.
In Section~\ref{sec:spectra} we present analysis of the spectroscopy and comparison with other objects.
The earliest spectrum is modelled in Section~\ref{sec:modelling}, along with light curve modelling to test varying scenarios.
In Section~\ref{sec:discussion} we introduce and describe a self-consistent interpretation of the progenitor system and in Section~\ref{sec:conclusions}, we summarise the main conclusions of this work.

\section{Discovery, classification and data collection}\label{sec:data}
\sn2018gjx was discovered by the Xingming Observatory Sky Survey (XOSS) on UTC 2018-09-15 19:41:24 (MJD 58376.82) at 16.7~mag in a clear filter.
The Asteroid Terrestrial-impact Last Alert System survey \citep[ATLAS;][]{Tonry2018,KWSmith2020} recorded a non-detection of ATLAS-$c$ $>20.2$~mag on MJD 58375.6.
We take the estimated explosion date as half-way between the two, MJD 58376.2, which is likely accurate to within a day (see Section~\ref{sec:lcmod}).
The transient was originally classified by the extended Public ESO Spectroscopic Survey of Transient Objects \citep[ePESSTO;][]{Smartt2015} as a young SN II \citep{classification}.

The ATLAS-$o$ and $c$ band images were processed as described in
\cite{Tonry2018} and photometrically and astrometrically
calibrated immediately \citep[using the RefCat2 catalogue][]{Tonry2018b}. Deep reference images are subtracted from each ATLAS image during the night, and transient
objects are catalogued on these difference images as described in
\cite{KWSmith2020}. In automated discovery mode,
point-spread-function photometry is carried out on the difference
images and all sources greater than 5$\sigma$ are recorded.
For all real astrophysical transients, ATLAS performs forced
photometry at the mean position of the detected transient in all
recent images after the reference has been subtracted. For SN\,2018gjx, 
point-spread-function fitting was forced at the position of the object
and all magnitudes reported here are in the AB system.
Typically ATLAS will observe with a quad of $4\times30$\,sec per
night and we have calculated weighted averages of these on a nightly
basis.

Photometric follow-up was obtained as part of the Las Cumbres Observatory \citep[LCO;][]{Brown2013} Global Supernova Project \citep[GSP;][]{Howell2017} using the Sinistro cameras mounted on the LCO network of 1\,m telescopes.
Spectroscopy was also obtained using the Floyds spectrographs on the LCO 2\,m telescopes at the Siding Spring and Haleakala observatories. These spectra were reduced using the LCO Floyds pipeline. 
Imaging and spectroscopy were also obtained with the 2\,m Liverpool Telescope \citep[LT;][]{Steele2004} using IO:O and the Spectrograph for the Rapid Acquisition of Transients \citep[SPRAT;][]{Piascik2014}, and the Floyds spectrographs on the 2\,m LCO telescopes at Haleakala and Siding Spring observatories.
Photometry was performed on the LT and LCO frames using a custom {\sc python} routine and {\sc pyraf} as part of {\sc astroconda}, the instrumental magnitudes were calibrated to Sloan Digital Sky Survey \citep[SDSS;][]{Ahn2014} stars in the field.

Spectroscopy was also obtained by ePESSTO using the ESO Faint Object Spectrograph and Camera (v.2) \citep[EFOSC2;][]{Buzzoni1984} mounted on the 3.58\,m New Technology Telescope (NTT) based at the European Southern Observatory's (ESO) La Silla observatory in Chile.
NTT Spectroscopic data reduction and calibration was done through standard pipelines\footnote{https://github.com/svalenti/pessto}. 
A final spectrum was obtained with Keck+LRIS on 2019-02-05 6:20:1 (MJD 58519.26) and was reduced the standard way using the LPIPE pipeline\footnote{https://www.astro.caltech.edu/~dperley/programs/lpipe.html}. 
Spectra were also taken with the Nordic Optical Telescope (NOT), equipped with ALFOSC, and the Telescopio Nazionale Galileo (TNG) with DOLORES. These spectra were reduced in a standard way, consisting of a wavelength calibration against an arc lamp and a flux calibration utilising a spectrophotometric standard star. Additional difference image photometry was obtained through the Zwicky Transient Facility \citep[ZTF;][]{Bellm2019,Graham2019} public stream through the Lasair broker\footnote{https://lasair.roe.ac.uk/} \citep{KWSmith2019}.

\begin{table}
    \centering
    \caption{The properties of \sn2018gjx and its environment.}
    \begin{tabular}{ll}
    \hline
     RA (J2000) & 02:16:15.63 \\
     $\delta$ (J2000) & +28:35:20.40 \\
     Host & NGC 865   \\
     $z_\mathrm{host}$ & $0.010 $    \\
     $z_\mathrm{SN}$ & $0.012  $    \\
     $\mu$ &  $32.7\pm{0.3}$ mag \\
     $D_\mathrm{L}$ & $35\pm{5}$ Mpc \\
     Non-detection (MJD) & 58375.6 \\
     Discovery (MJD) & 58376.82\\
     Estimated explosion date (MJD) & 58376.2 \\
     \Emw & 0.078 mag  \\
     \Eh & $0.12\pm{0.03}$ mag \\

     \hline
    \end{tabular}
    \label{tab:props}
\end{table}

\subsection{Host properties and distance}
\begin{figure}
    \centering
    \includegraphics[scale=0.52]{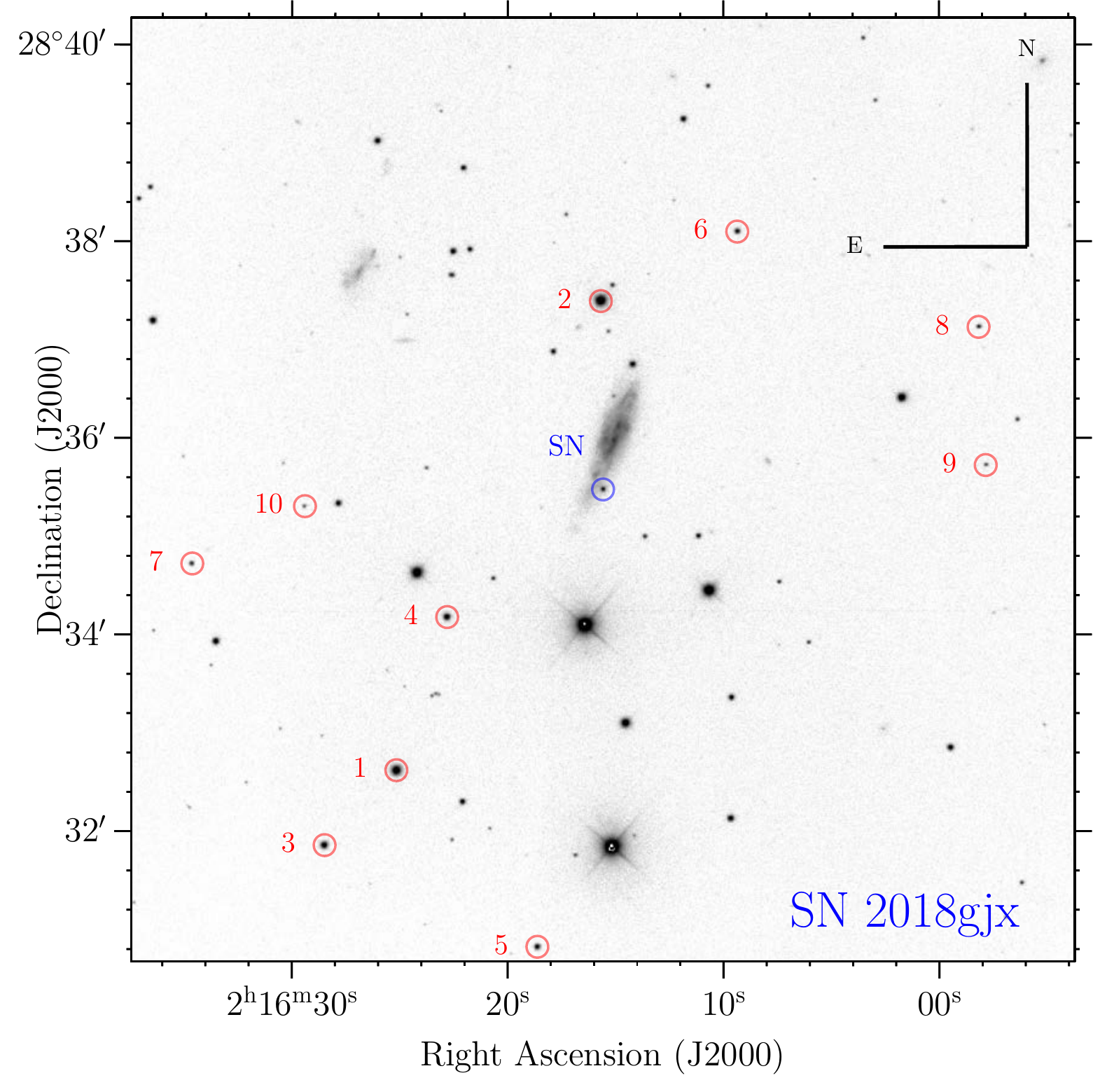}
    \caption{Liverpool Telescope $r$ band image of \sn2018gjx (blue circle) in NGC 865 and 10 stars used to calibrate the photometry are labelled as red circles. }
    \label{fig:host}
\end{figure}

Figure~\ref{fig:host} shows the location of \sn2018gjx in NGC 865 at 30.4 arcsec from the centre of the host.
NGC 865 is a spiral galaxy of type Sbc with signs of possible irregularity or interaction. It is highly inclined, likely to a minimum of 80 degrees as per the axial ratio of 0.2 \citep[see HyperLEDA\footnote{http://leda.univ-lyon1.fr/};][]{Makarov2014}.
The redshift reports of NGC 865 in the NASA/IPAC Extragalactic Database\footnote{http://ned.ipac.caltech.edu/} (NED) range from $z=0.0099-0.012$.
The NTT classification spectrum of \sn2018gjx displays H Balmer P-Cygni lines consistent with $z=0.012$; these lines are broader than typical galaxy emission lines and so appear to be due to the transient.
A spectrum of the host galaxy near the transient was obtained from one of our SPRAT observations and was consistent with $z=0.012$, while a SPRAT observation of the core of NGC 865 agrees with $z=0.010$. We take these values as the redshift of the transient and the host respectively.
Given that the event occurred in the outskirts of an inclined spiral galaxy, this discrepancy could be attributed to the difference in rotation within the galaxy. Radio measurements of NGC 865 give the maximum rotational velocity as $\sim 300$ \kms\ \citep{Haynes2018}, which is half the velocity difference between the galaxy centre and the transient location, and itself exceeds the maximum rotational velocity measured by \cite{Galbany2014} for a sample of SN hosts.
A more exotic explanation could be that the transient is in an indistinct dwarf galaxy, merging with the southern portion of NGC 865.

We use cosmological parameters from the Nine-year Wilkinson Microwave Anisotropy, $H_0=69.32$ km s$^{-1}$ Mpc $^{-1}$, $\Omega_m=0.286$, $\Omega_\Lambda=0.714$ \citep{Hinshaw2013}. 
The median distance modulus of 16 Tully-Fisher measurements given in NED is $\mu=32.7$ mag, with standard deviation of 0.3 mag. The most recent value is $\mu=32.52\pm{0.45}$ mag \citep{Tully2016} and is consistent with the median. 
Here we adopt the mean values of $\mu=32.7\pm{0.3}$ mag as the distance modulus, which gives a luminosity distance $D_\mathrm{L}=35\pm{5}$ Mpc, and a projected distance of 5.2 kpc/deprojected distance of $r_\mathrm{SN}= 8\pm{2}$ kpc from the centre of the host to the transient. 

Finally, the absolute $B$ band magnitude of NGC 865 is $-20.5\pm{0.5}$ mag. Using the relationship between $M_B$ and oxygen abundance from \citet{Tremonti2004}, a ratio of $r_\mathrm{SN}/R_{25} \approx 1$, and an average metallicity gradient of $\sim -0.47$  $r_\mathrm{SN}/R_{25}$ \citep{Pilyugin2004,Tartaglia2020}  gives a rough estimate of $12+ \mathrm{log(O/H)}\sim 8.56\pm{0.15}$ dex.
Comparison with $12+ \mathrm{log(O/H)}\sim 8.69\pm{0.05}$ dex for Solar metallicity \citep{Asplund2009}
suggests around Solar to sub-Solar metallicity at the location of the transient. 
In contrast, the absence of N~{\sc ii} in the spectrum extracted from near the SN (which points to ionisation from star formation), while the absence of H$\beta$/O~{\sc iii} suggests Solar or greater metallicity. Which could point to star formation in the region due to ongoing interaction with the host and a companion as previously suggested.

\subsubsection{Extinction}

The dust maps of \citet{Schlafly2011} give a Galactic reddening in the direction of NGC 865 of \Emw\ $=0.078$ mag. Since \sn2018gjx exploded close to a dust lane on the outer edges of a  high inclination galaxy, it may be expected that the host reddening \Eh\ is non-negligible.
Examination of the spectra at the position of the \NaI\ D lines shows no strong and consistent absorption at significant S/N, suggesting that \Eh\ is relatively small, but signatures of interaction can mask this line. 
Comparison of the continuum shape of the first spectrum with other hot and blue objects gives an estimate of \Eh\ $\sim 0.1$ mag.
Modelling of this spectrum, see Section~\ref{sec:specmod} finds consistency with this value at \Eh\ $= 0.1-0.15$ mag.
Thus, we adopt \Eh\ $=0.12\pm{0.03}$ mag, giving \Etot\ $= 0.20\pm{0.03}$ mag and corrected for reddening using the extinction law of \citet{CCM}.
The general properties of the host and transient that we use are given in Table~\ref{tab:props}.

\section{light curves} \label{sec:lc}
\begin{figure}
    \centering
    \includegraphics[scale=0.52]{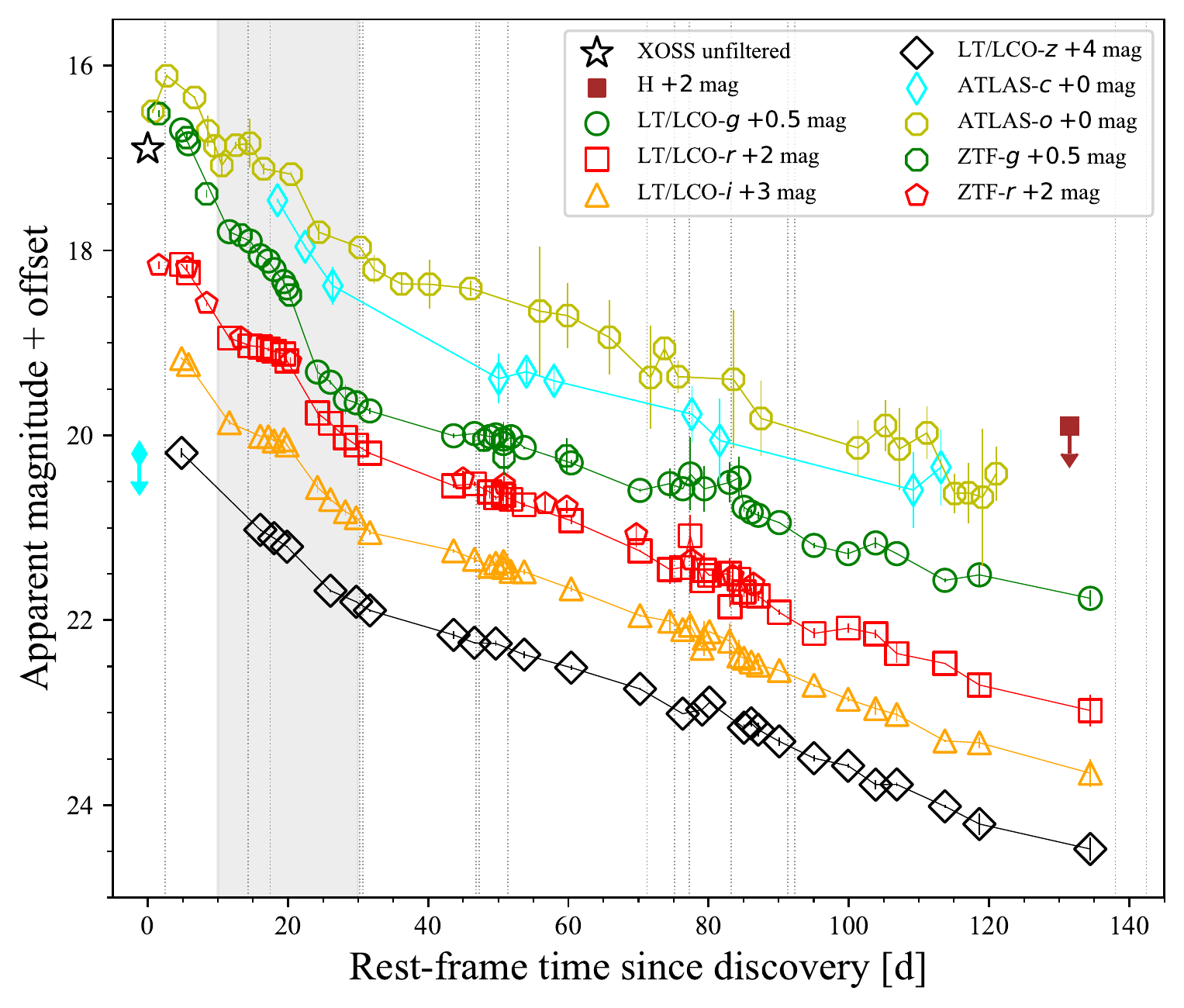}
    
    \caption{Multi-colour light curves of \sn2018gjx from the LT, LCO, and ATLAS, as well as public ZTF $g$ and $r$ data and the XOSS clear filter detection. The ATLAS $c$ non-detection is shown as a solid cyan diamonds with downward-facing arrow, while the $H$ band non-detection at 130 d is shown as dark red square with downward-facing arrow. The grey vertical dashed lines represent epochs of spectral observations. The grey band denotes the approximate epochs of the non-interacting broad SN component dominated SN IIb phase, after this the spectra are dominated by interaction signatures.}
    \label{fig:LCs}
\end{figure}

The multi-colour light curves of \sn2018gjx are shown in Fig.~\ref{fig:LCs} and presented in Table~\ref{tab:phot}. 
The $o$ band is the only band which shows the rise to peak of approximately $ 3.9$ days from the last non-detection.
After maximum light the multi-colour bands are better observed, and from this point they decline for $\sim 10$ days before levelling off.
This flattening of the light curve is seen most prominently in the $o$, $r$, and $i$ bands, and less so in the $g$ and $z$ bands, although the latter is poorly sampled around this time.
If one considers an early luminous and short-lived component in addition to a typical SE-SN light curve, where the bluer bands peak earlier, then such evolution is explained.
This phase lasts for approximately 10 days before returning to a rapid decline for a further 5 days. 
The period when the flattening occurs and its subsequent rapid decline (from 10 - 30 d since discovery), coincides with the phase when the spectra of \sn2018gjx look most like a SN IIb (with broad SN components and no obvious signatures of CSM interaction, see Section~\ref{sec:snphase}.)
The light curves then evolve onto a slower decline between days 40 -- 80, with average decay rates in $g$, $r$, $o$, $i$, and $z$ at day 50 of 0.020, 0.024, 0.025, 0.024, and 0.024 mag d$^{-1}$ respectively.
The light curve slope steepens after this and we see an average linear decline at 100 days of 0.021, 0.025, 0.028, 0.029 and 0.029 mag d$^{-1}$ for $g$, $r$, $o$, $i$, and $z$, respectively.

H-poor interacting SNe can be brighter in the NIR than in the optical at 100 days, this was seen most prominently in Type Ibn SN 2006jc \citep{Pastorello2007}.
This has been attributed to dust formation when accompanied by a rapid decline in the optical light curves or asymmetries in emission lines \citep[see,][]{Pastorello2008b,Bevan2020}, or from the heating of pre-existing dust \citep[`infrared echo'; e.g.,][]{Tartaglia2020}.
We observed \sn2018gjx in the $H$ band using IO:I on the LT some 130 days after first detection.
Both \sn2018gjx and \sn2006jc show interaction with He-rich CSM in their spectra (see Section~\ref{sec:spectra}),  if \sn2018gjx was similar in brightness to \sn2006jc in $H$, then the scaled $H$ magnitude would be $\sim 17.3$ mag.
A 538~s exposure was obtained by dithering 9 separate 59.8~s exposures.
From this, we were able to derive an upper limit of $m_H>17.9$ mag ($M_\mathrm{H}>-14.8$ mag).
Although this limit is marginally deeper than the estimated luminosity in the $H$ band of \sn2006jc, it is insufficiently deep to rule out a contribution to the luminosity by thermal emission from dust.

\begin{table*}
    \centering
    \caption{Table describing the photometric measurements of SN 2018gjx.}
    \begin{tabular}{lcccccccc}
    \hline
     MJD     & $g$ & $r$ & $i$ & $z$ & $c$ &$o $ & Facility \\
        & [mag]      &     [mag]    &   [mag]    & [mag]      &   [mag]    & [mag] & \\
     \hline
      58375.61 &   -    &      -       &   -    &   -    & $>20.2$      & -
         & ATLAS \\
     58377.58  &   -    &      -       &   -    &   -    & -     & 16.49$\pm{0.02}$
         & ATLAS \\
    58381.70  & 16.20$\pm{0.02}$ &	16.15$\pm{0.02}$ &	16.17$\pm{0.02}$ &	- & - & -
         & LCO \\   
    58382.69 & 16.35$\pm{0.02}$ &	16.24$\pm{0.02}$ &	16.23$\pm{0.02}$ & - & - & -& LCO \\   
     \hline
     \multicolumn{8}{l}{The full table can be found in machine readable format online.}
    \end{tabular}
    
    \label{tab:phot}
\end{table*}

\subsection{$g-r$ colour curves}\label{sec:cc}
\begin{figure}
    \centering
    \includegraphics[scale=0.54]{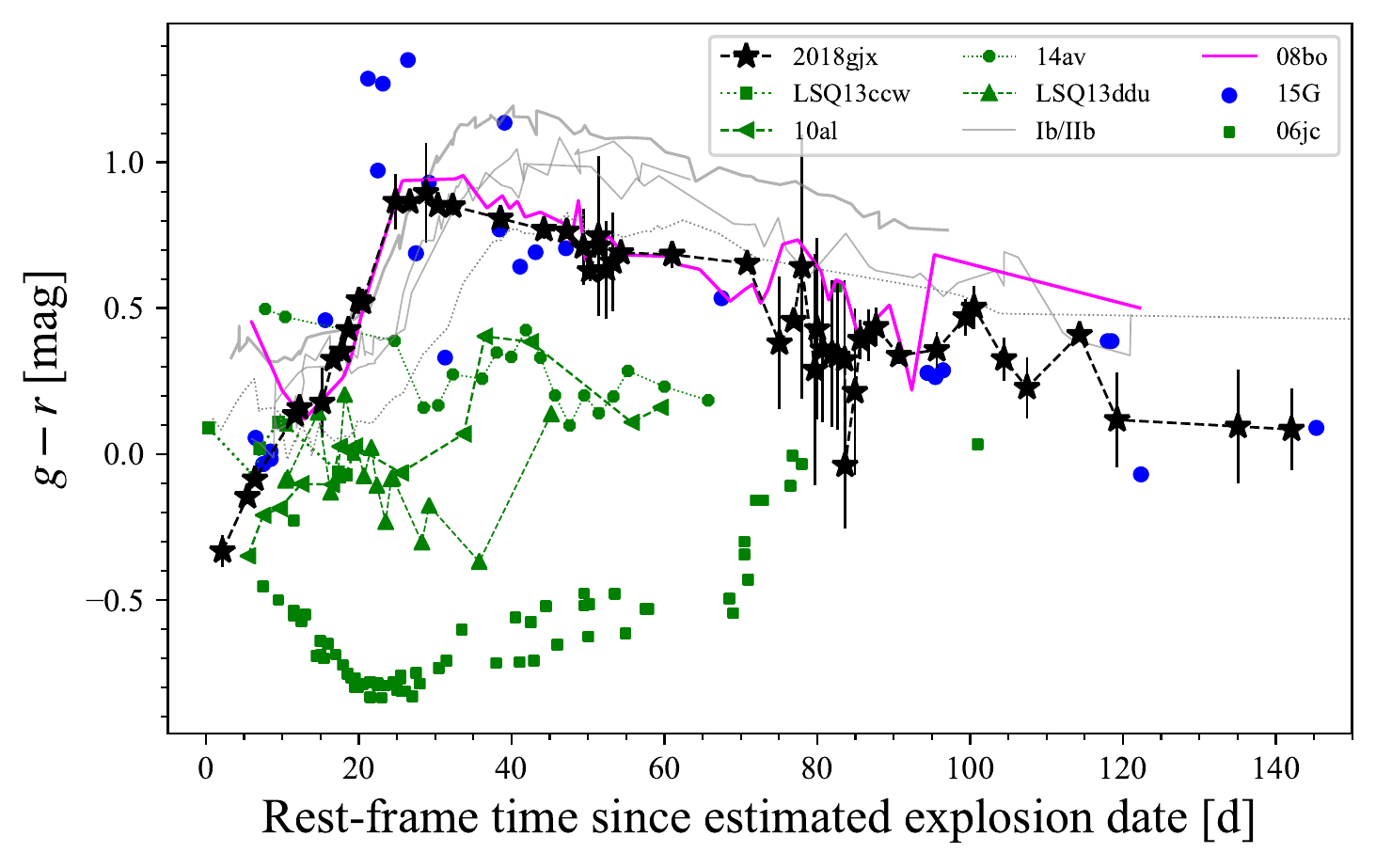}
    \caption{The $g-r$ colour curve of \sn2018gjx compared with a selection of SNe Ib/IIb \citep[grey, see][]{Prentice2016}, and a sample of SNe Ibn; SN 2006jc \citep{Foley2007,Pastorello2007}, SN 2014av \citep{Pastorello2016}, SN 2015G \citep{Shivvers2017b}, SN 2010al \citep{Pastorello2015c}, LSQ13ddu \citep{Clark2020}, and LSQ13ccw \citep{2015MNRAS.449.1954P}. SNe 2006jc, 2015G, and 2010al were not observed in $g-r$, the colours are derived from synthetic photometry obtained from the flux-calibrated spectra. The Type Ibn SN 2015G (blue dots) and Type IIb SN 2008bo (purple line) are emphasised owing to some similarity in their photometric and spectroscopic characteristics to \sn2018gjx. }
    \label{fig:ccs}
\end{figure}

The $g-r$ colour curves of SN\,2018gjx are presented in Fig.~\ref{fig:ccs}, and compared with the colours of some normal SNe IIb and SNe Ibn. 
The change in filter usage (e.g.~from \textit{UBVRI} to \textit{ugriz}) over the last decade means that not all objects were observed in the same bands. This is especially relevant to SNe Ibn where very few observations are found to have been made in $g$ and $r$.
To correct for this, the $g-r$ colour curves of SN\,2006jc, SN\,2010al, and SN 2015G are constructed by extracting synthetic $g$ and $r$ magnitudes from their flux-calibrated spectra.

Multi-colour photometry of SN\,2018gjx was first obtained a few days after explosion. From this epoch, the $g-r$ colour curve is defined by a redward evolution for $\sim25$ days, followed by a blueward evolution. 
The transition occurs at the same time as light curve evolves onto the linear tail.
Similar evolution, across a similar timescale, is seen in Type IIb SN 2008bo \citep{Bianco2014}, with the sole difference over the first 120 days being that SN 2008bo has an initial blueward evolution until $\sim$ 12 days after explosion that is not seen in SN\,2018gjx.
This blue/red/blue evolution is commonly seen in SE-SNe \citep{Prentice2016}.
However, SNe with strong shock-cooling tails in their light curves \citep[e.g., SNe 1993J, 2013df, 2016gkg;][]{Richmond1994,MG2014,Tartaglia2017} evolve straight from blue to red at these early times, and it is to these that SN 2018gjx is most similar.
Comparatively, SNe Ibn show great diversity in their colour evolution.
SN 2015G is most like SN 2018gjx and is plotted using the estimated explosion date of three days prior to the discovery date \citep[2015-03-23.778, but see][for a discussion on \texp]{Shivvers2017b}.
SN 2006jc displays an inverted evolution to that of SN 2018gjx, and two other SNe Ibn LSQ13ddu \citep{Clark2020} and SN 2014av \citep{Pastorello2016}, appear to follow the trend of blue to red, albeit from a bluer starting point.

\subsection{Pseudo-bolometric and absolute $r$ band light curves}
\begin{figure}
    \centering
    \includegraphics[scale=0.57]{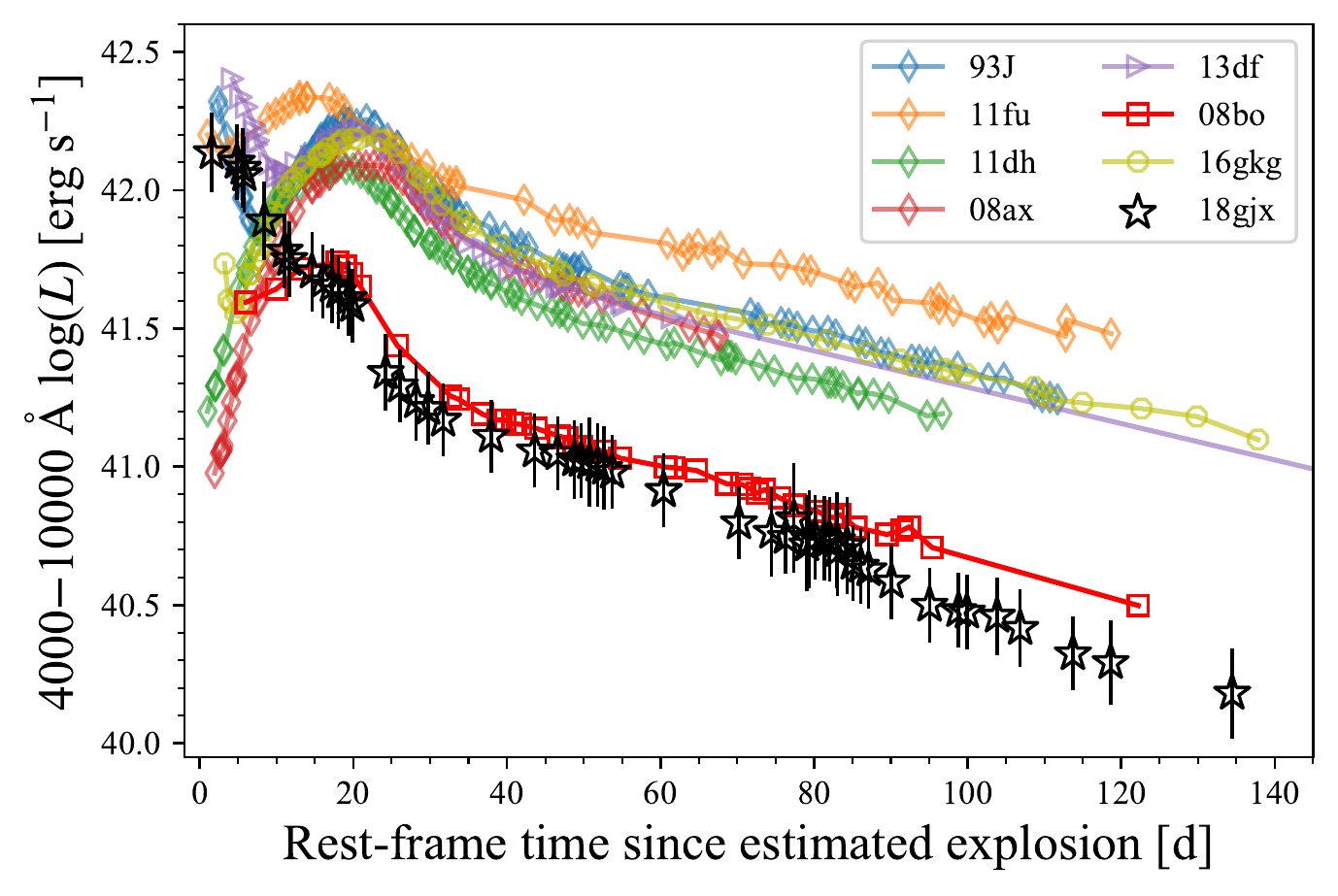}
    \includegraphics[scale=0.57]{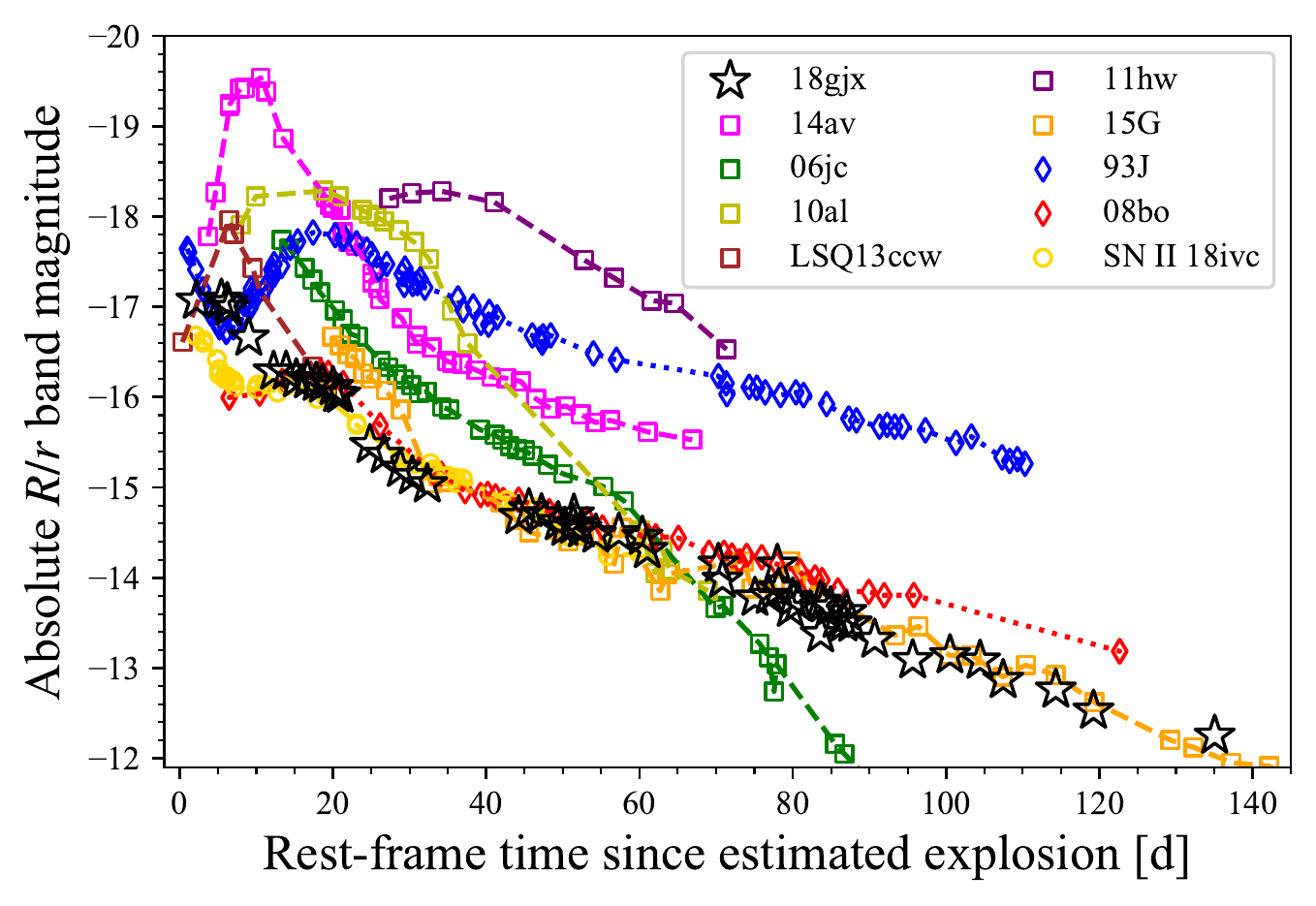}
    \caption{ {\bf Upper panel:} The pseudo-bolometric light curve of \sn2018gjx compared with SNe IIb. The unusual SN IIb 2008bo has a remarkably similar light curve after 10 days. {\bf Lower panel:}  Absolute $R/r$ LCs of SNe Ibn (squares) and SNe IIb (diamonds) in comparison with \sn2018gjx (black stars). Type Ibn \sn2015G (orange), Type IIb \sn2008bo (red), and Type IIb \sn2018ivc (gold)show similar evolution. The dramatic decline of \sn2006jc is attributed to dust formation. Data sources not already cited in text: SN 2011fu \citep{2015MNRAS.454...95M}, SN 2011dh \citep{Marion2014}, SN 2008ax \citep{Pastorello2008}, SN 2011hw \citep{Pastorello2015c}.
    }
    \label{fig:bol}
\end{figure}

An optical pseudo-bolometic light curve was constructed by converting the dereddened multi-colour photometry to flux and building spectral energy distributions (SEDs) for each date. These SEDs were then integrated over $4000-10000$ \AA, covering the effective range of the $griz$ bandpasses, and converted to pseudo-bolometric luminosity using a fixed luminosity distance of 35 Mpc. This method is described in more detail in \cite{Prentice2016}. 
The pseudo-bolometric light curve of \sn2018gjx is shown in Fig.~\ref{fig:bol} in comparison with SNe IIb and SNe Ibn.
It is a requirement of the bolometric light curve methods that the transients are observed in $BVRI$, $gri$, or equivalent and while some interpolation and extrapolation can be done to account for missing data we find that many SNe Ibn are only well-sampled in one or two bands, which limits our comparison sample.
The peak luminosity of $\sim 1.4\times 10^{42}$ \ergs\ places the first peak in a similar range to that of the SNe IIb that show shock-cooling tails. After this the light curve evolution diverges and fails to rise a second peak. It instead tracks the evolution of \sn2008bo, although with a steeper linear decline of $\sim 0.026$ mag d$^{-1}$. The decline is steeper than for most SNe IIb \citep{Prentice2019}

For further comparison, the absolute $r$ band light curve of SN\,2018gjx along with those of a sample of SNe IIb and SNe Ibn, with a focus on Type Ibn, is also shown in Fig.~\ref{fig:bol}. 
Also included in this plot is the unusual interacting Type II SN 2018ivc \citep{Bostroem2020}, which has an almost identical $r$ light curve to \sn2018gjx but evolves to show H-rich interaction spectra.
Observations in the $r$ band are unavailable for some events so the $R$ band is used instead. Both bandpasses probe similar regions of the spectrum so the shapes of the light curves are comparable but result in a marginally greater luminosity in $R$ of 0.1 -- 0.2 mag \citep[estimated via the colour conversions of][]{Jordi2006}, 
with a peak $r \sim $ --17.1 mag, \sn2018gjx stands out as considerably less luminous than the other objects except for the low-luminosity SN IIb 2008bo and Type Ibn \sn2015G. 
As can be seen in Fig.~\ref{fig:bol}, the $R-$band light curve shape of \sn2015G is very similar to \sn2018gjx $r$ band, with a similar flattening around 20 days and similar later decay rate. Unfortunately, it is unknown if \sn2015G had a similar first peak to \sn2018gjx, but the similarity in light curves between the two objects suggests that \citet{Shivvers2017b} were accurate in their estimated explosion date of 20 days before discovery.

\section{Spectroscopy of SN 2018gjx}\label{sec:spectra}
\begin{figure*}
    \centering
    \includegraphics[scale=0.87]{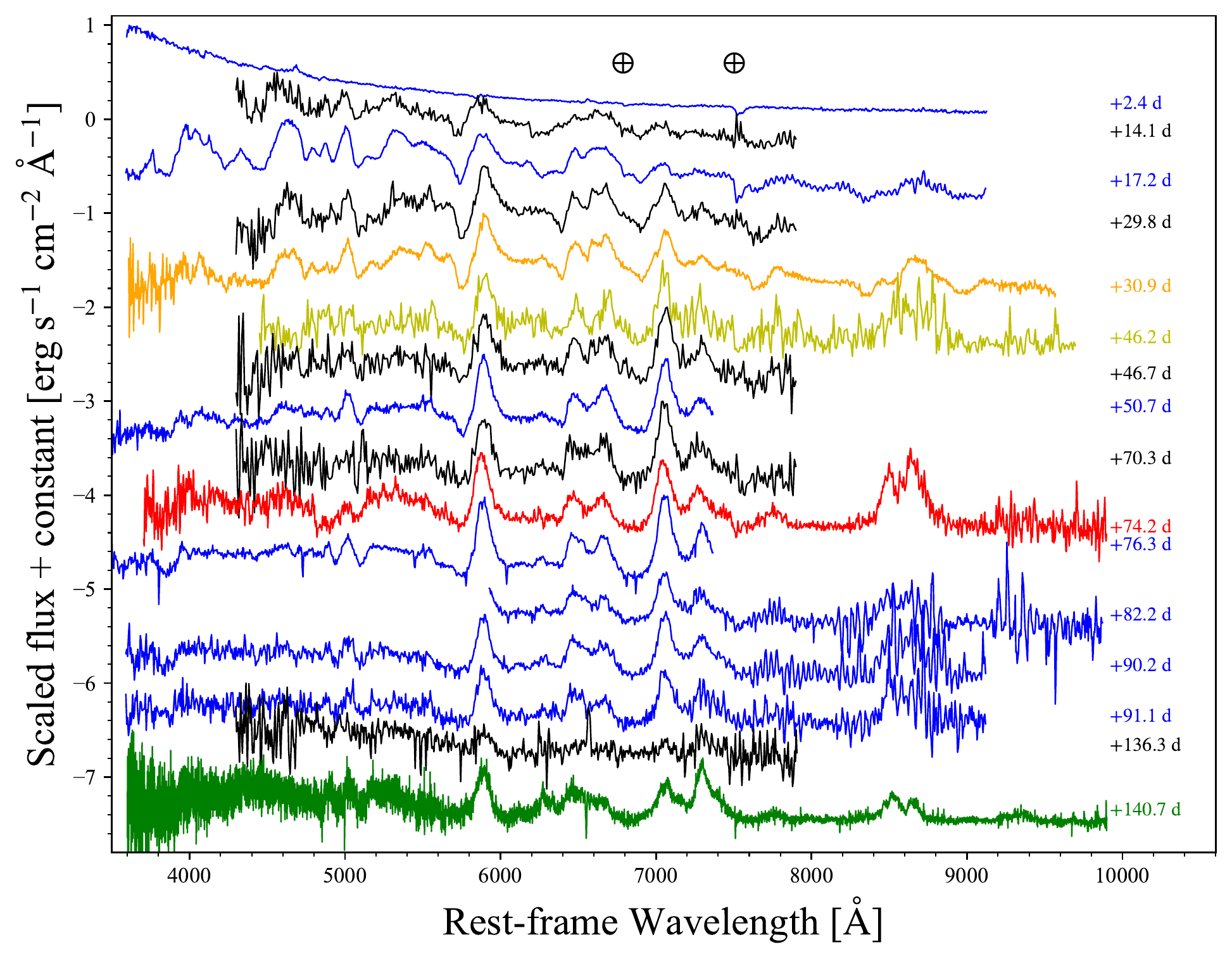}
    \caption{Spectra of SN 2018gjx from NTT (blue), LT (black), NOT (Orange), LCO (yellow), TNG (red), and Keck (green). Epochs are rest-frame $t-t_\mathrm{discovery}$ while the positions of strong Telluric absorption is denoted by the ``Earth'' symbols. The initial spectrum is hot and blue, with signatures of UV-ionised CSM lines (see Fig.~\ref{fig:earlyspec}). Ten days later the transient resembles a Type IIb SN with strongly evolving \HeI\ lines, and by 47 days after discovery the SN spectrum has given way to an interaction spectrum dominated by strong \HeI\ lines. }
    \label{fig:spectra}
\end{figure*}

The spectra of SN 2018gjx cover from 2 to 140 days after discovery and are shown in Fig.~\ref{fig:spectra}.
The spectral evolution of the object is unusual, passing through three distinct phases; a hot, blue phase (Phase I), a a non-interacting broad SN component dominated SN IIb phase (Phase II), and a He-rich interaction phase, similar to SN Ibn (Phase III).
In this section we investigate each phase in turn.

\subsection{Phase I: The hot, blue phase} \label{sec:earlyspec}

\begin{figure*}
    \centering
    \includegraphics[scale=0.75]{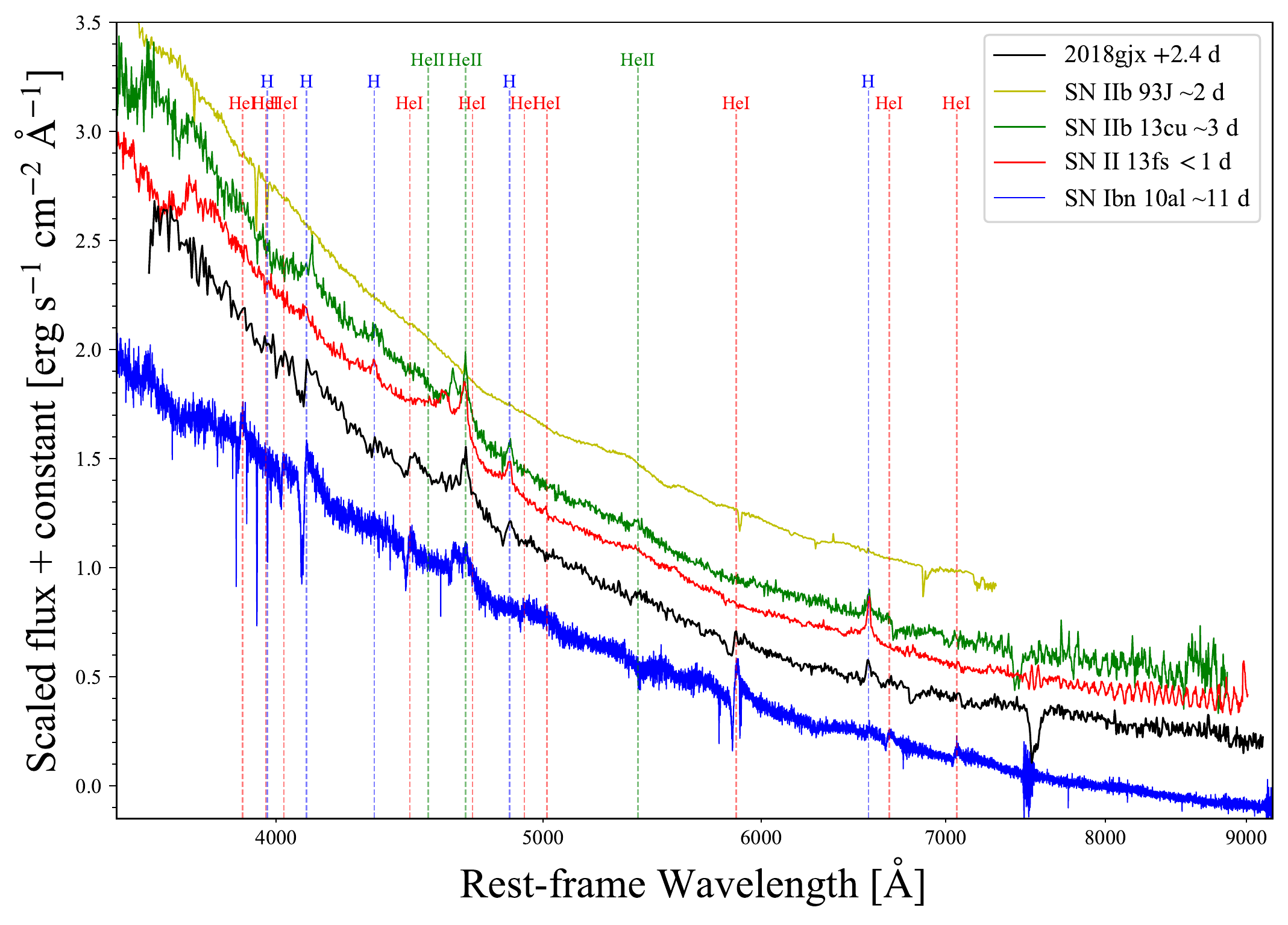}
    \caption{The $+2.4d$ spectrum of SN 2018gjx compared with other SN spectra which have UV-ionised CSM features, all phases are relative to discovery \citep[Data sources;][]{Matheson2000b,Pastorello2015c,Galyam2014,Yaron2017}. There is similarity between SN 2018gjx and SN Ibn 2010al in terms of H Balmer P-Cygni features in the spectrum. }
    \label{fig:earlyspec}
\end{figure*}

UV photons from the shock-breakout (SBO) or SN ejecta/CSM interaction can lead to strong but short-lived emission lines in unshocked CSM, particularly of \HeII\ \lam 4686 \citep{Chugai2001,Galyam2014,Kochanek2019}. These are seen in very early spectra of a number of SN classes; SNe IIb, \citep[e.g., \sn1993J and \sn2013cu;][]{Benetti1994,Matheson2000,Galyam2014}. 
SNe II \citep[See,][]{Niemela1985,Smith2015b,Yaron2017,2018ApJ...861...63H,2016ApJ...818....3K}, plus interacting SNe of Type IIn 1998S \citep{Shivvers2015} and Type Ibn 2010al \citep{Pastorello2015c}. 
Comparison with some of these objects are shown in Fig.~\ref{fig:earlyspec}.

The earliest spectrum of SN 2018gjx was taken 2.4 days after discovery and coincides with the $o$ maximum. 
The spectrum at this time is hot and blue, a black-body fit to this spectrum gives a temperature of $19\,000 - 21\,000$ K, with the caveat that the peak of the SED is not observed. There are weak P-Cygni features attributable to the H-Balmer series and \HeI.
The \HeII\ \lam 4686 emission feature is also present and has a broad base and a narrow peak, its presence is indicative of ionised CSM.
These features are evidence of nearby CSM being present at the time of explosion. 

The ZTF SEDM spectrum\footnote{https://wis-tns.weizmann.ac.il/object/2018gjx} taken 23 hours later (MJD 58380.24) does not display the prominent \HeII\ feature, but does display the weak \HeI\ and H features as seen in the $+2.4$ d spectrum, which suggests that either the ionised He in that region has completely recombined,  or that the shock has reached its position and that this represents the edge of the CSM \citep{Yaron2017,Kochanek2019}. However, we note that the SEDM is of lower spectral resolution and makes identification and comparison of narrow features more difficult, it is sufficient to detail the \HeII\ feature \citep{Bruch2020}.

\subsubsection{P-Cygni line velocities at +2.4 days}
The absorption minima of the various P-Cygni lines is $\sim 1500$ \kms, see Table~\ref{tab:linevels}.
A P-Cygni feature at 5410 \AA\ could be \HeII\ \lam 5412 with a velocity of $\sim 1800$ \kms, which is a similar to the other P-Cygni lines.
The expansion velocities given by these features suggest the outflow from a hot, massive star, as they are not high enough to be SN ejecta but are much greater than the few tens to hundred \kms\ seen in red supergiant winds.
The wind velocities of Wolf-Rayet stars have been measured to be a few hundred to a few thousand \kms\ \citep{Crowther2007,Sander2012}. 
Outburst from Luminous Blue Variables (LBV) can also reach velocities of several thousand \kms, and LBV stars transitioning to a Wolf-Rayet phase \citep{Smith2020} have been suggested as progenitors for some SNe Ibn \citep{Smith2012}. These SNe typically show weak narrow \Ha\ emission in their interaction spectra \citep{Pastorello2008b}, and while such features are absent in the later spectra of \sn2018gjx (see Section~\ref{sec:IbnPhase}), the presence of narrow H features during Phase I and broad H features during Phase II (see Section \ref{sec:snphase}) suggests that the progenitor star was not entirely H-deficient.

\begin{table}
    \centering
    \caption{Line velocities in the +2.4 d spectrum, $\dag$ is ambiguous identification. }
    \begin{tabular}{ccc}
     
    \hline
     Rest-$\lambda$ & Line I.D.    & Velocity, $v$ \\
     
     [\AA]   &   & [\kms] \\
     \hline
     4102 & H$\delta$ & $1500\pm{600}$ \\
     4341$^{\dag}$ & H$\gamma$ &   $1300\pm{600}$      \\
     4472 & \HeI & $1000\pm{500}$ \\ 
     4861 & H$\beta$ & $1600\pm{500}$   \\
     5412 & \HeII &  $1800\pm{500}$    \\
     5876 & \HeI &  $1600\pm{500}$     \\
     6563  & H$\alpha$  & $1600\pm{500}$ \\
     6678$^{\dag}$ & \HeI &  $1500\pm{700}$     \\
     7065$^{\dag}$ & \HeI &  $1400\pm{500}$     \\
    \hline     
    \end{tabular}
    \label{tab:linevels}
\end{table}

\subsection{Phase II: The SN IIb phase}\label{sec:snphase}
The next spectrum, taken at $+14.1$ d, is very different to the  spectrum at $+2.4$ days and is dominated by broad P-Cygni profiles.  A prominent P-Cygni feature consistent with \HeI\ \lam5876 is seen with an absorption minimum at $8000$ \kms. At this velocity, other absorption features in the spectrum line up with strong \HeI\ lines of 5016, 6678, and 7065 \AA. 
A bump at $\sim 6563$ \AA\ suggests the emergence of a broad \Ha\ emission. 
\citet{Rigault2019} classified the transient as a SN IIb on the basis of a $\sim$ 30 day ZTF SEDM spectrum. This is consistent with the results of {\sc SNID} \citep{Blondin2007} and Gelato\footnote{https://gelato.tng.iac.es/gelato/} \citep{Harutyunyan2008} that find top matches of our spectra at this phase to SNe IIb.

\begin{figure*}
    \centering
    \includegraphics[scale=0.76]{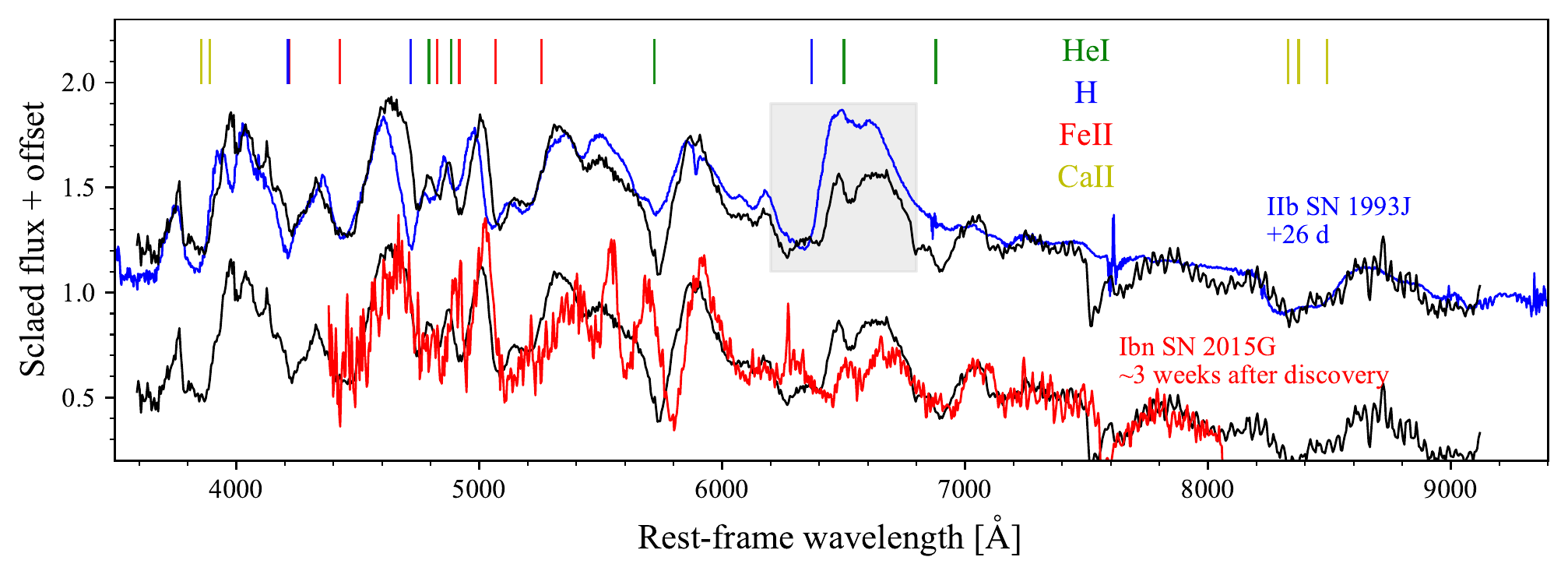}
    \caption{Comparison of the SN 2018gjx spectrum (black) during the Phase II ($+17.2$ d) with Type IIb \sn1993J (blue) at maximum light and Type Ibn \sn2015G (red) during its brief photospheric phase. Also shown are the positions of strong lines of \FeII\, H, \HeI\, and \CaII\ corresponding with absorption minima. \sn2018gjx shares many similarities with both of the comparison SNe, but a key area is around \Ha\ (grey shaded region), where the transient displays a P-Cygni feature similar to \sn1993J and other SNe IIb. The equivalent feature in \sn2015G is weak and short-lived. \HeI\ is stronger in \sn2018gjx at this time compared with SNe IIb however, which suggests that we are seeing some component of interaction, as per \sn2015G. }
    \label{fig:snphase}
\end{figure*}

 In Fig.~\ref{fig:snphase} we compare the $+17.2$ d spectrum with that of Type IIb SN 1993J \citep{Matheson2000} and Type Ibn SN 2015G \citep{Shivvers2017}.
Both \sn1993J and \sn2015G have features in common with \sn2018gjx. The feature around \Ha\ is stronger in \sn1993J than in \sn2018gjx, which is in turn stronger than in \sn2015G. \sn2015G lacks a prominent absorption component to this feature, which is seen in the other two objects. 
The visible absorption features between 3500 \AA\ and 6500 \AA\ are mostly similar in the three objects if one accounts for a slight velocity shift. It is apparent however, that \sn1993J and \sn2018gjx are most similar in location and relative strength of these features. 
An important difference is in the strength of the \HeI/\NaI\ D P-Cygni feature around 5890 \AA\ and the strength of \HeI\ \lam7065, which are most similar between SN 2018gjx and \sn2015G. These lines strengthen over time and this suggests that we are seeing signs of interaction with He-rich CSM at this epoch. 
The velocity of the absorption component, and of the other visible \HeI\ lines at 6678 \AA\ and 7065 \AA, better matches \sn1993J. From $+10$-- $+30$ d we consider the spectra of \sn2018gjx to be most similar to the non-interacting SNe IIb, and in its spectroscopically-defined `Phase II'.

\subsubsection{Phase II line velocities}
\begin{figure}
    \centering
    \includegraphics[scale=0.7]{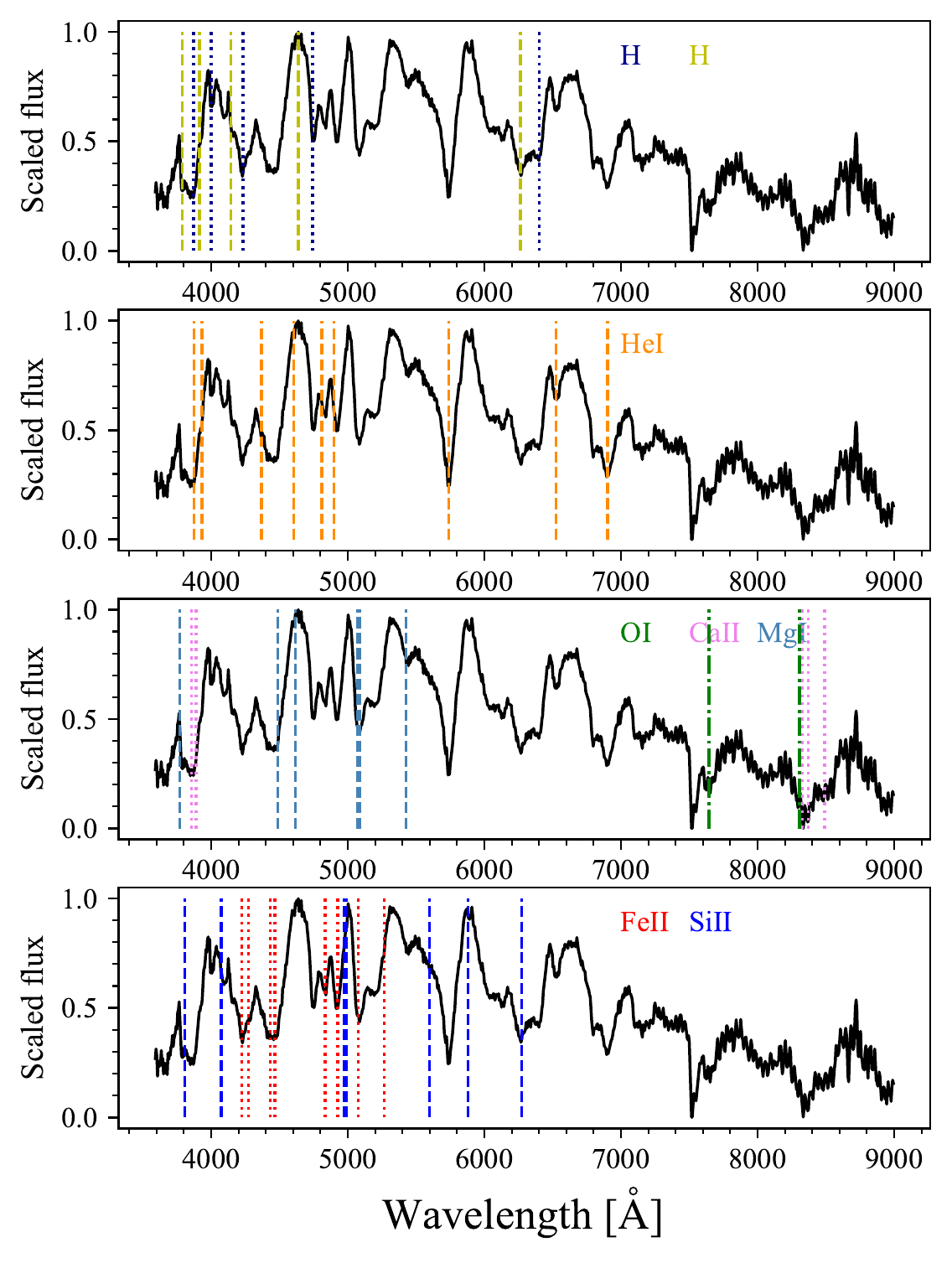}
    \caption{The $+17.2$ d spectrum, in the SN IIb phase, where the spectra show broad SN components and no obvious signs of CSM interaction. The line positions for various ions at the velocities described in the text. (Top) High/low hydrogen (yellow/blue), (upper middle) \HeI, (lower middle) \OI, \MgI, \CaII, (lower) \SiII, \FeII. }
    \label{fig:IIbvels}
\end{figure}

Figure~\ref{fig:IIbvels} shows the $+17.2$ d spectrum with the positions of various line velocities marked.
The \HeI\ lines are significantly stronger than in the previous spectrum at $+14.1$ d, with velocities of $7000$ \kms\ as measured from the absorption minima, and velocities at the edge of the blue wing of 11000--12000 \kms. 
\Ha\ is visible as a P-Cygni feature with both absorption and emission components. The emission component shows a ``notch'', indicative of absorption from \HeI\ \lam6678. The absorption profile has two minima, the high-velocity component has a velocity of 14000 \kms, while the low-velocity component is at 8000 \kms. 
The remaining Balmer lines from H$\beta - \epsilon$ are visible, with absorption velocities consistent with the He-component values.
If \SiII\ is present, then it could contribute to the \Ha\ ``high velocity'' absorption component, this would give a velocity of 4000 \kms, consistent with other measured velocities. It may also account for a feature around 4770 \AA.
Ambiguity about the presence of \SiII\ in the optical spectra of SNe IIb has been discussed in the literature \citep{Prentice2017}.
Oxygen is identified through the \OI\ \lam7774 line, a velocity of 5000 \kms\ can be measured from the minimum blueward of this. The low velocity of this line separates it from the telluric {\sc O$_2$} absorption.
Calcium is present as both the H\&K lines and the near infra-red triplet, with a velocity of $6000$ \kms\ measured for each line.
There are plausible matches for ions of iron-group elements, \FeII\ and \MgI, both at $5000$ \kms.
The measured line velocities are entirely consistent with those of SNe IIb \citep{Prentice2017,Prentice2019}.

The final spectrum in this phase of evolution was taken at $+29.8$ d.
The \HeI\ P-Cygni and \Ha\ features are now significantly stronger than previously seen. The blue component of the \Ha\ absorption is no longer present, the extent of the absorption is at 11000--12000 \kms, as previously measured for \HeI\ \lam5876.


\subsection{Phase III: The Type Ibn phase}\label{sec:IbnPhase}

During Phase III, epochs after $\sim 40$ days, the spectra are dominated by prominent emission features of \HeI, of which the strongest is \HeI\ \lam5876, see Fig.~\ref{fig:spectra}. This line is scattered into the \NaI\ D line as is seen in some SNe Ibc and Ibn \citep{Pastorello2015a,Pastorello2015b, Jerkstrand2015}. Also visible are the  \CaII\ NIR triplet and an emission line at $\sim 7300$ \AA, which is interpreted as [\CaII] \lam\lam 7324, 7292 in SNe IIb \citep[e.g.][]{Jerkstrand2015} and as \HeI\ \lam 7281 in SNe Ibn \citep{Pastorello2016}.
Two strong emission features around 5000 \AA\ are seen to evolve with time. This could be due to two \HeI\ lines at 4922 \AA\ and 5016 \AA\ or two \FeII\ lines at 4924 \AA\ and 5018 \AA. \citet{Jerkstrand2015} attributed the emission in the $202$ d model spectrum of Type IIb \sn2011dh to \HeI\ \lam 5016, but also found contribution from \FeII\ in other models. These two lines are also seen in the early nebular phase of He-poor SNe, which suggests they may predominantly consist of \FeII, rather than \HeI. Since the identification of these lines is ambiguous, we treat them as a likely blend of both \FeII\ and \HeI\ and omit them from further analysis.

\begin{figure}
    \centering
    \includegraphics[scale=0.6]{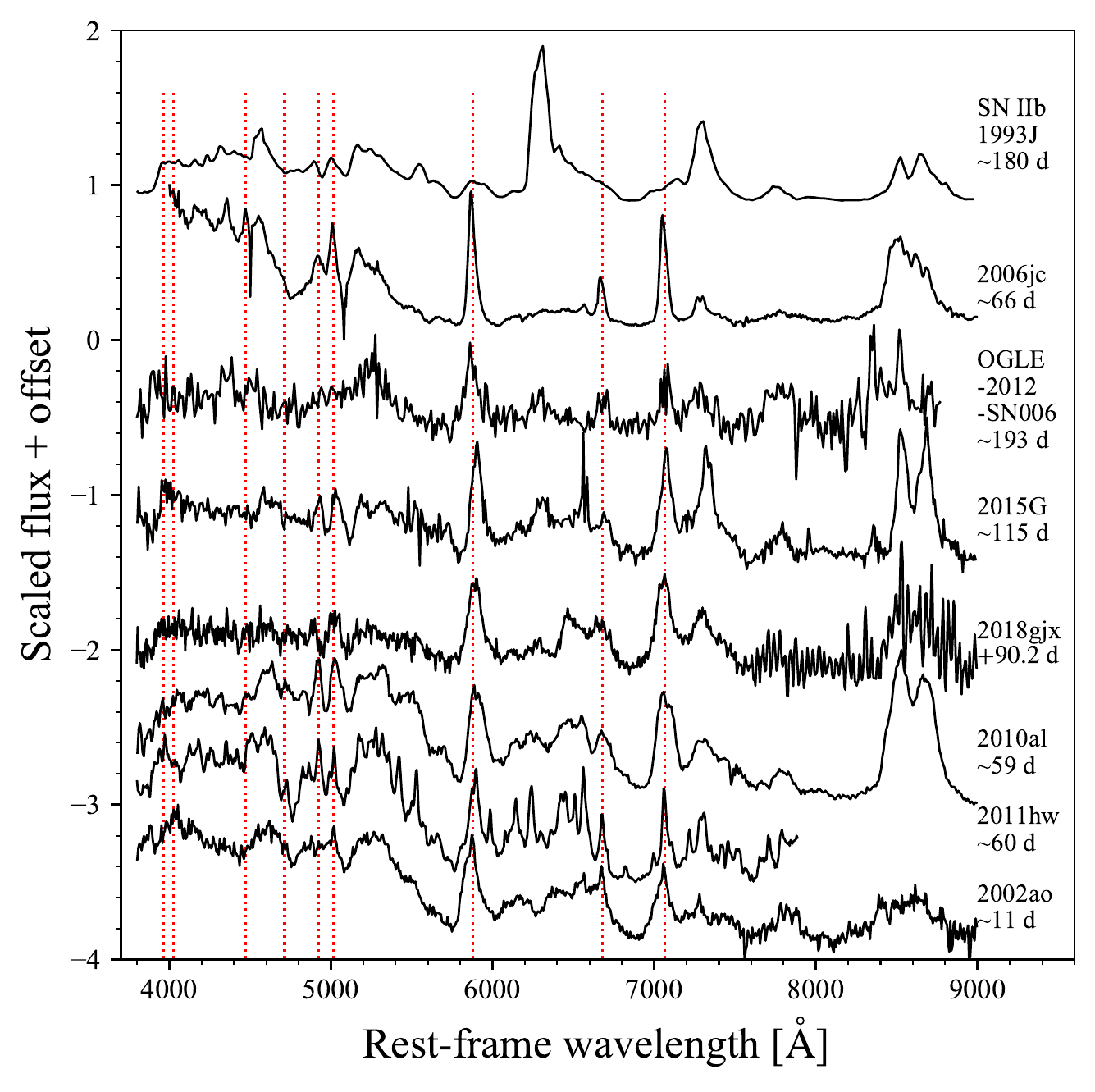}
    \caption{The $+90.2$ d spectrum of \sn2018gjx is shown compared to the spectra of a sample of SNe Ibn, where very similar spectroscopic features are seen, in particular the strong emission profiles of \HeI\ lines (red dashed lines denote their rest wavelengths). These spectra are at varying phases relative to discovery and cover canonical Type Ibn like \sn2006jc, as well as those that display a conventional SN-like photospheric phase \citep[see][]{Hoss2017}. Also for comparison is the spectrum of Type IIb SN 1993J during the nebular phase, demonstrating a lack of similarity with this SN type, in particular it lacks the strong \HeI\ lines seen in \sn2018gjx. Data sources additional to those already given: OGLE-2012-SN06 \citep{2015MNRAS.449.1941P}, SN 2002ao \citep{Pastorello2008b}.   }
    \label{fig:isibn}
\end{figure}

Figure~\ref{fig:isibn} shows how the $+90.2$ d spectrum of \sn2018gjx compares with the spectra of SNe Ibn at various phases, as well as the canonical Type IIb \sn1993J during the nebular phase \citep{Matheson2001}. The spectrum of \sn2018gjx does not show the strong blue emission from the forest of Fe lines commonly seen in SNe Ibn but it does show the characteristic \HeI\ emission profiles that define this SN-type and are signatures of CSM interaction. This late time spectrum of \sn2018gjx is also unlike the spectrum of \sn1993J; \sn1993J does not show the emission components of \HeI, which demonstrates  that despite a similar appearance to SNe IIb during Phase II, it does not show a similar spectroscopic evolution into the later epochs.

\subsubsection{Characterising emission features using Gaussian line fitting}
\begin{figure*}
    \centering
    \includegraphics[scale=0.8]{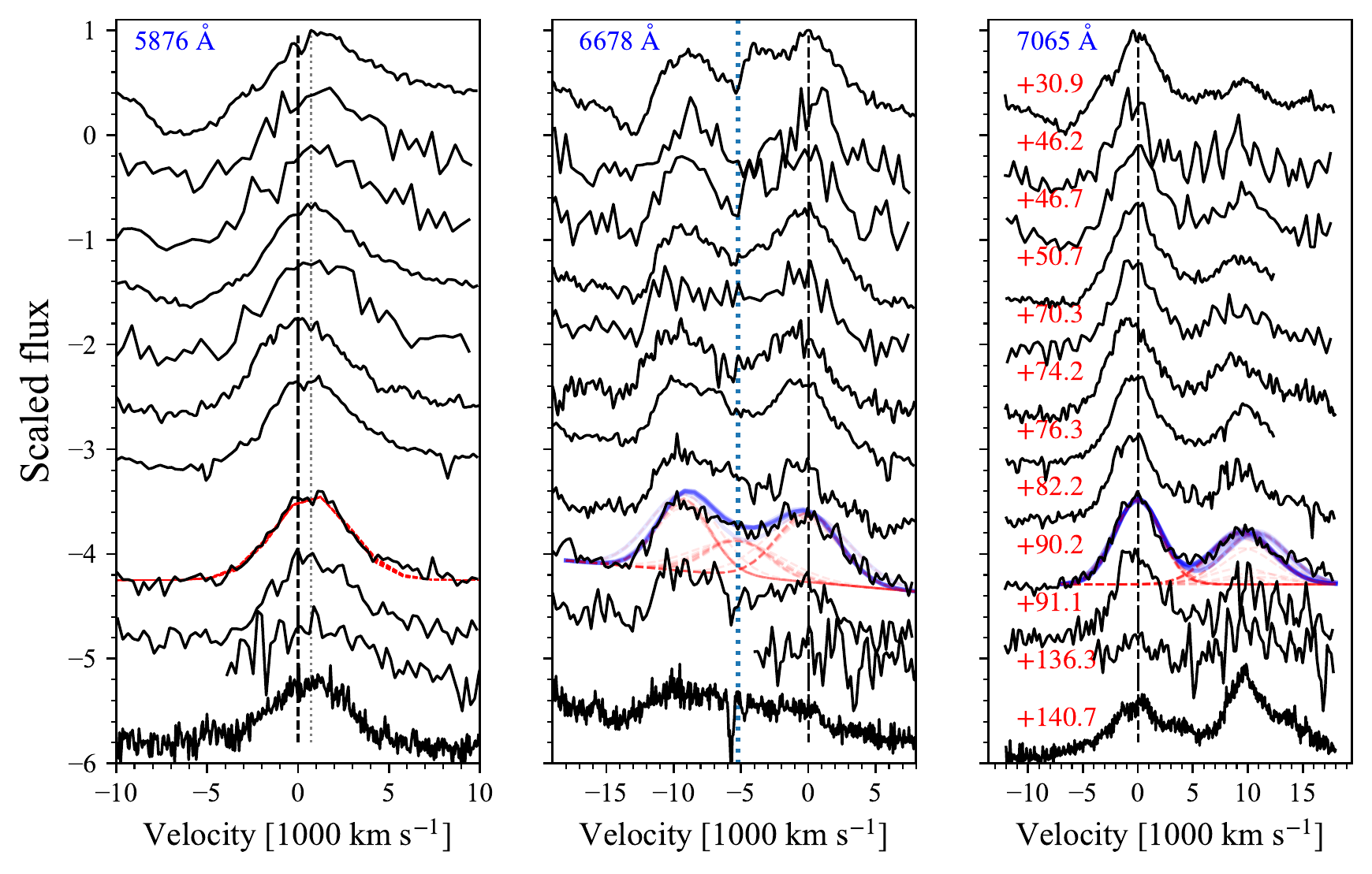}
    \caption{The evolution of emission lines centred on 5876, 6678 and 7065 \AA. The dashed red lines on day $+90.2$ demonstrate the Gaussian line fitting in red for each component used and in blue for the combined model. {\bf Left panel:} Clearly shown is that the emission feature near 5876 \AA\ (dashed line) is centred around 5890 \AA\ (dotted line), which shows re-emission of the \HeI\ \lam5876 flux through the \NaI\ D line, as has been suggested for other SNe Ibn. {\bf Centre panel:} The 6678 \AA\ feature is blended with another feature blueward of it, by $+136.3$ d this line is no longer visible and we see instead a broad flat-topped profile centred on \Ha, which is denoted by the blue dotted line. {\bf Right panel:} \HeI\ \lam7065 is the strongest and most unambiguous of the \HeI\ lines. It is blended with other lines, including [\CaII] \lam\lam7292,7324 on the redward side.}
    \label{fig:Heseq}
\end{figure*}

Figure~\ref{fig:Heseq} shows the evolution of the emission lines around 5876, 6678, and 7065 \AA\ from $+46.2$ d until $+140.7$ d. The emission line at 5890 \AA\ is mostly likely the  \HeI\ \lam5876 scattered into the \NaI\ D line \citep{Pastorello2015a,Pastorello2015b, Jerkstrand2015}. Despite this, we will continue to refer to it as \HeI\ \lam5876 as it is emission from this line transition that dominates the flux.
There remain signs of absorption to the blue of this feature until about $\sim90$ days, with $v\approx 6000$ \kms\ as measured from the absorption minimum.
This line remains symmetrical throughout, with little indication of a flux excess at either wing, nor is there evidence of any blueshift in the lines, which was seen in \sn2006jc by 100 days and was attributed to dust-formation \citep{Smith2008,Chugai2009}.

The feature at 6678 \AA\ is consistent with \HeI\ \lam 6678 line blended with an emission feature to the blue (see Fig.~\ref{fig:spectra}) which evolves from the broad \Ha\ emission seen during the SN~IIb phase. A possible absorption component is seen to the blue at a velocity of $v\approx 5\,000-6\,000$ \kms, which is consistent with that seen for the \HeI\ \lam 5876 absorption.
Over the course of 100 days the \HeI\ \lam 6678 line loses definition and by the $+140.7$ d spectrum it ceases to be a distinct feature. 

The \HeI\ \lam 7065 line is also shown in Fig.~\ref{fig:Heseq} and is blended with a broad emission line of [\CaII]. The Keck spectrum at $140.7$ days is of higher resolution than previous spectra and the shape of this feature suggests there are more than just two components. Other potential features in this region include [\FeII] \lam7155 and \HeI\ \lam7281 but they are typically far weaker.

To isolate and extract the parameters of the three potential \HeI\ emission lines shown in Fig.~\ref{fig:Heseq} around 5876, 6678, and 7065 \AA\, we have fitted a series of Gaussian profiles to the data in velocity space from $+50$ days (demonstrated for the He lines in Fig.~\ref{fig:Heseq} for the $+90$.2 d spectrum). The procedure is as follows, firstly a linear continuum is set between two points either side of the features. The continuum reference points were averaged over a small window to account for noise variations and allowed to vary in position to provide some estimate of uncertainty on the continuum itself.  The central wavelength and width of the individual Gaussians was allowed to vary, the former to account for velocity offsets. The parameters extracted from the fits were the peak, width, and centre, which were used to calculate the characteristic velocity in the form of the Full Width Half Maximum (FWHM) and the integrated flux.

In the fitting of each region, a number of potential weaker features were included. For the 5876 \AA, along with \HeI\ 5876 \AA, we included components corresponding to \NaI\ D doublet. For the 6678 \AA\ region, we included three components, corresponding to \HeI\ 6678 \AA, \Ha\ 6563 \AA, as well as an additional unidentified component to the blue of \Ha\ in an attempt to better replicate the shape of the emission around this region. The effect of this is to force the \HeI\ feature to match the redward emission  of this region, which provides the best constraint for this emission line. The fitting process is most uncertain for this \HeI\ 6678 \AA\ line, and results in large uncertainties in the measured FWHM.

For the 7065 \AA\ region, we used a single \HeI\ \lam7065 and [\CaII] \lam\lam 7272, 7324 components due to the limits of the spectroscopic resolution and signal-to-noise. The exception to this is the late Keck spectrum which we fit with these two components, plus a further two [\CaII] \lam\lam 7272, 7324 components, [\FeII] \lam 7155, and \HeI\ 7281.  For the region around [\CaII], it was seen that the emission could be described by a series of broad and narrow components at varying velocity offsets, these may or may not be physical but provide a representation of the total flux. It was also found that the \HeI\ 7281 feature could not be included with the constraint that the FWHM matched that of the \HeI\ \lam 7065 line and so it was disregarded.

\begin{figure}
    \centering
    \includegraphics[scale=0.7]{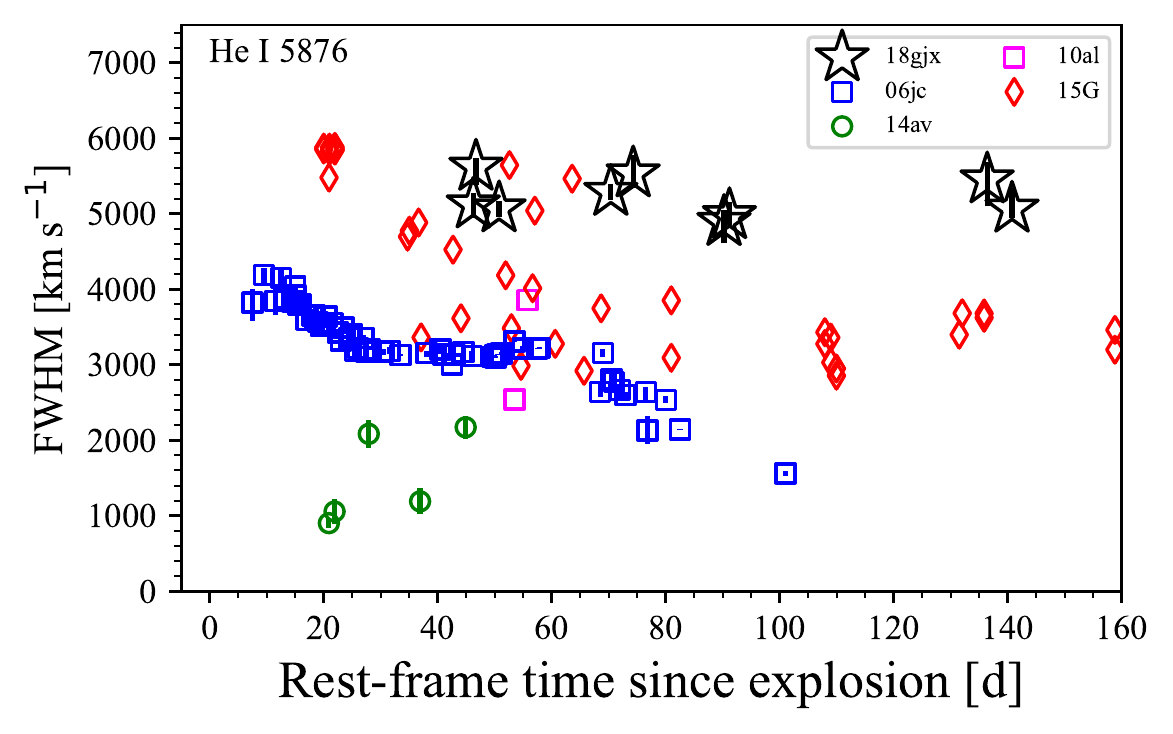}
    \includegraphics[scale=0.7]{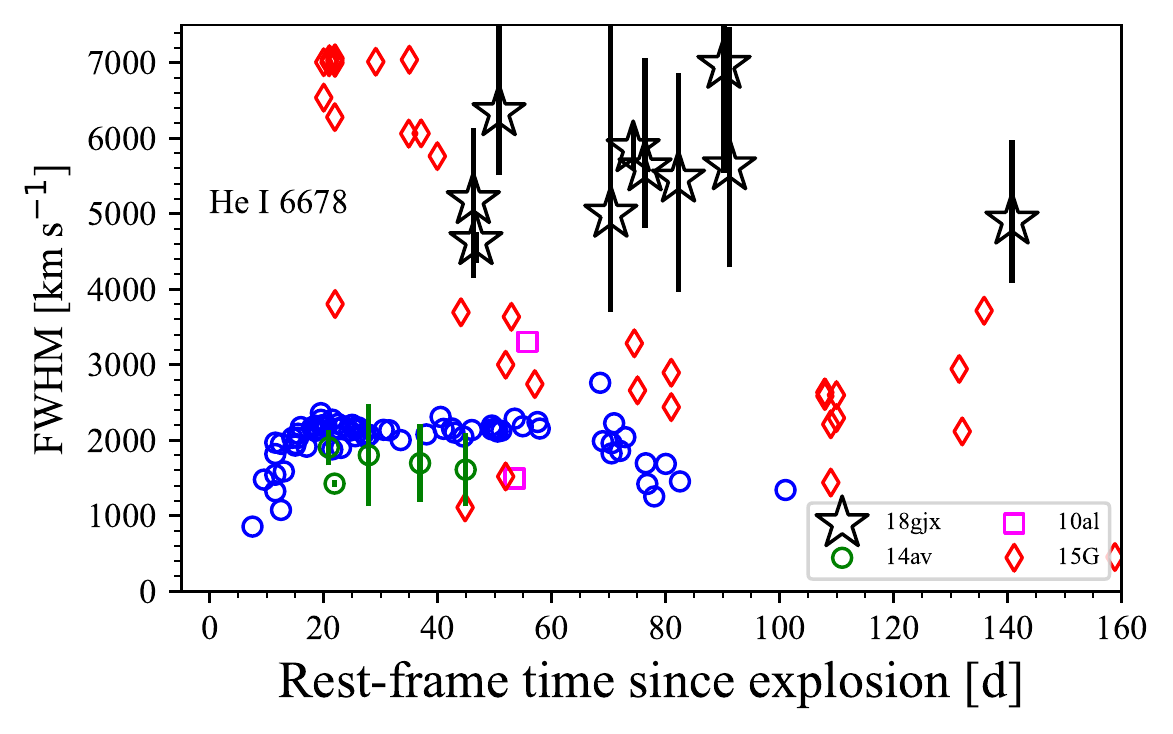}
    \includegraphics[scale=0.7]{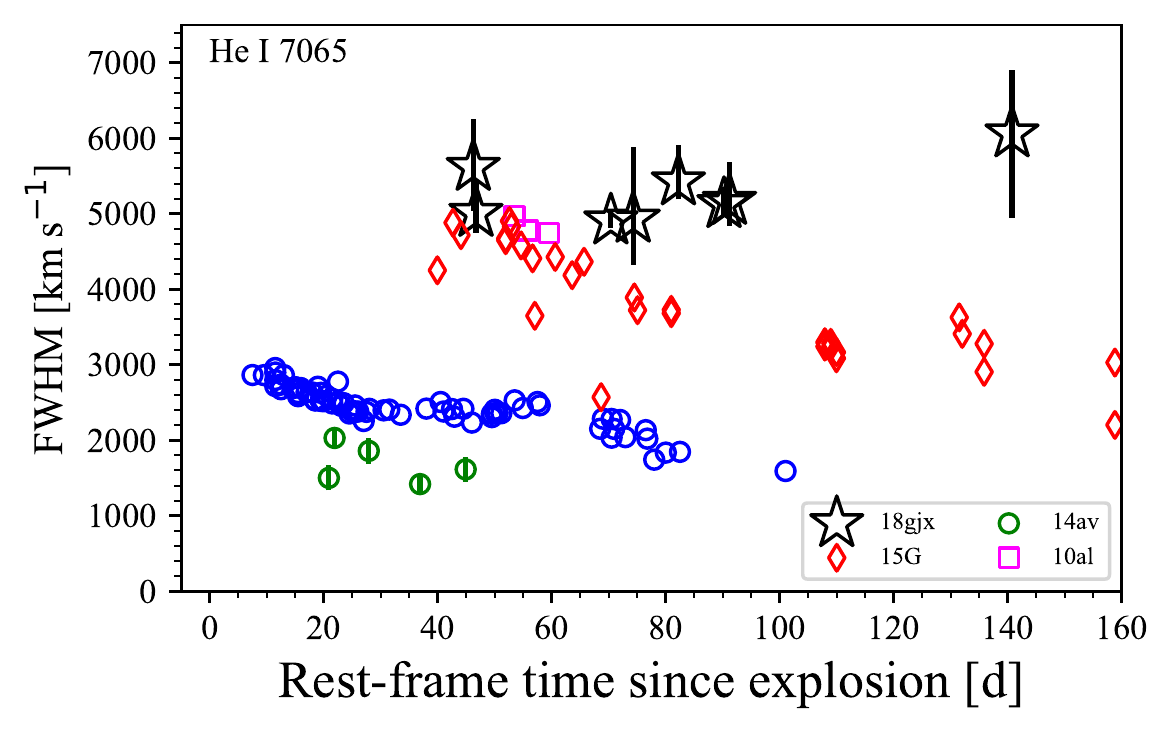}
    \caption{Comparison of intermediate-width \HeI\ emission line FWHM velocities for \sn2018gjx and SNe Ibn. The measured velocities remain greater than the comparison objects and show no discernible evolution over time. }
    \label{fig:FWHMs}
\end{figure}

The FWHM measurements for the three \HeI\ lines are compared with a small sample of SNe Ibn in Fig.~\ref{fig:FWHMs}. 
There exist few SNe of this type with spectroscopic observations beyond a couple of months, which limits the sample size. The FWHM is calculated in the same way as \sn2018gjx for \sn2006jc, \sn2015G, and \sn2010al. Although we find that the observed \HeI\ emission lines in SN 2018gjx are well fit by single Gaussian profiles, both narrow and intermediate components can be seen in the emission lines of some SNe Ibn. \cite{Pastorello2016} fit multiple Gaussian components to the emission lines of SN 2014av to determine the properties of the narrow and intermediate components. We did the same for SN 2014av but found the fits and the offset of the weaker, broad component to be highly sensitive to the S/N and the choice of continuum, so we take the FWHM of the strongest component for comparison.  
We attempted to fit narrow and broad \HeI\ emission components for \sn2018gjx, but found that a single component \HeI\ emission is sufficient in the fitting.   
This could be intrinsic to the line profile or be a consequence of the resolution of the spectra. It can be seen that the emission lines of \sn2018gjx are considerably broader than for the other objects, with the closest being \sn2015G, and the FWHM does not appear to evolve in time.

The narrow components of the emission lines at late times in SNe Ibn typically have velocities in the range $800-1000$ \kms\ \citep{Pastorello2016}, with only three SNe in their sample having velocities below 600 \kms\ measurements (PS1-12sk, SN 2011hw and SN 2005al), and three substantially more than this (1900--2300 \kms). These narrow components measure the velocities of CSM outside the interaction region \citep{Pastorello2007}. Whilst we cannot distinguish a narrow component to the emission lines in \sn2018gjx, we do have an estimate of the velocity of material outside the photosphere in the early phases. As noted in Section~\ref{sec:earlyspec}, our $+2.4$~d spectrum has \HeI\ P-Cygni features consistent with velocities $\sim 1500$ \kms, which is at the upper end of the \cite{Pastorello2016} distribution. However, as discussed in \cite{Clark2020}, the resolution of the spectra is also important when making velocity comparisons and can result in higher measured values than are intrinsically present.

\subsection{The evolution of the 6563 \AA\ emission region}

\begin{figure}
    \centering
    \includegraphics[scale=0.8]{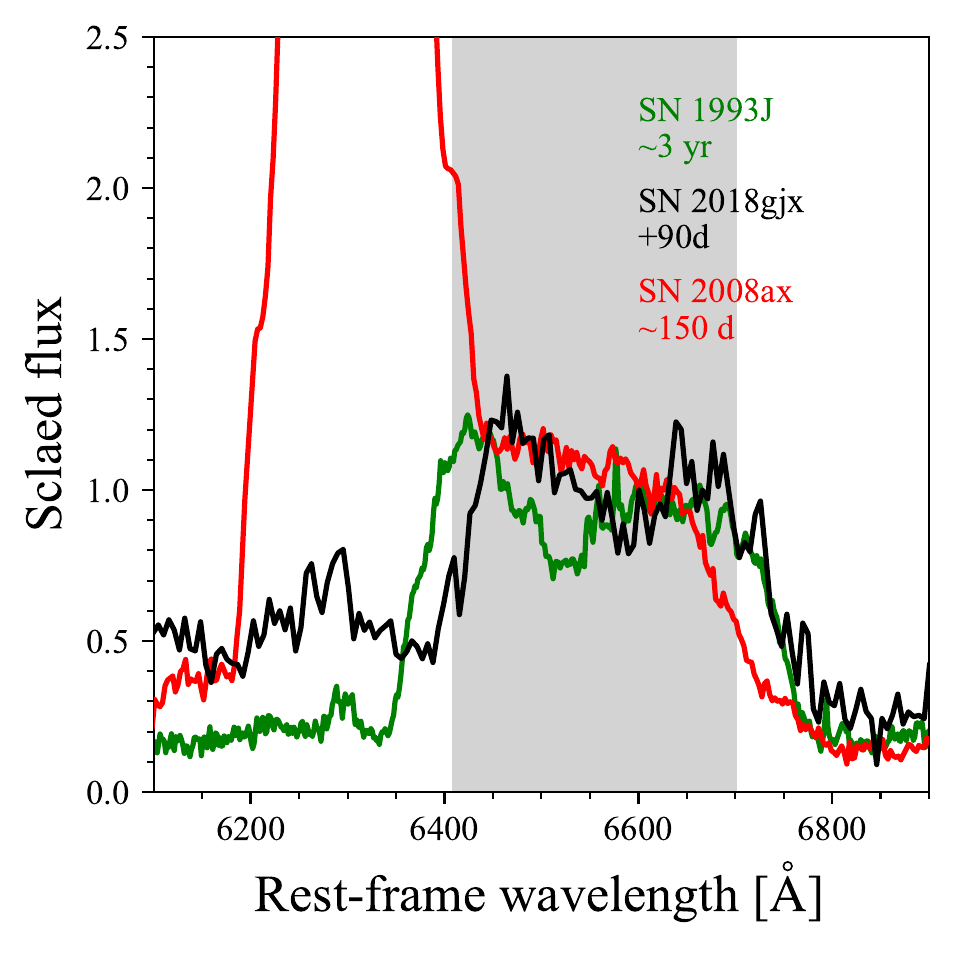}
     \caption{The $+90.2$ d spectrum of \sn2018gjx (black) against that of Type IIb \sn2008ax \citep{Taubenberger2011} in the nebular phase (red), which provides a good example of the flat-topped emission immediately redward of \Oneb\ \lam\lam 6300, 6364, and \sn1993J in the later interaction dominated phase \citep{Modjaz2014}. The spectra are scaled to the flux round 6600 \AA. The grey filled region is that identified by \citet{Jerkstrand2015} as N emission in the He layer. At this phase in \sn2018gjx, emission from \HeI\ \lam 6678 is extremely weak. }
    \label{fig:Hseq}
\end{figure}

During Phase II, there was a clear indication of the presence of the H Balmer series, with broad and strong \Ha\ emission. As the spectra transition into Phase III, the emission-dominated phase, this region becomes increasingly double-peaked, which may be due to a flat-topped profile with an absorption component centred on \Ha.  Figure~\ref{fig:Hseq} shows the spectrum of \sn2018gjx at $+$90.2 centred on the \Ha\ region. The profile has shoulders beginning at 5000 \kms\ and 6000 \kms\ with a width of 2000--3000 \kms. The total width of the emission profile changes little during the time of our observations. 

In the Phase II, the absorption component in this region would be consistent with coming from from \HeI\ \lam6678, which occupies the red wing of this emission feature. 
If this is also the cause here during Phase III, it requires that the \HeI\ line remains optically thick.
The red-wing of the broad flat emission feature becomes progressively weaker with time, reflecting the decreasing strength of the \HeI\ lines. While this occurs, the absorption component also weakens.
In SNe\,IIb, spectroscopic signatures of H weaken rapidly after peak and are absent during the nebular phase. In some examples of these events however, a broad flat shoulder is observed on the redward side of the \Oneb\ emission line at these phases. This has been interpreted as emission from [\NII] \lam \lam 6548, 6583 in the He shell \citep{Jerkstrand2015}, see Fig.~\ref{fig:Hseq}.
This interpretation is consistent with the properties of \sn2018gjx, e.g. the SN\,IIb-like nature of Phase II, the He-rich ejecta, and the high He and N surface abundance in the models of the early spectrum (see Section~\ref{sec:specmod}).

Alternatively, in the late-time spectra of H-rich CC-SNe, a flat-topped profile of \Ha\ is indicative of CSM interaction \citep[See,][]{Matheson2000,2017ApJ...834..118M,2017MNRAS.471.4047A,Chevalier2017,Weil2020}. This interaction usually takes place after many hundreds of days but with potentially significant CSM nearby to the progenitor of \sn2018gjx, these signatures may appear at earlier times. The velocity width of the feature centred on \Ha\ is consistent with that seen in late interacting H-rich SNe, as shown in Fig.~\ref{fig:Hseq} compared with a $\sim3$ yr spectrum of \sn1993J. That this \Ha\ feature is flat-topped, while the He lines are not, can be explained by considering that the H is confined to a shell in the outer ejecta (and likely mixed with He). It is unclear from our observations which of these two scenarios (emission from [\NII] \lam \lam 6548, 6583 in the He shell or interaction with CSM containing H) is the correct interpretation. 

\subsection{The presence of calcium and oxygen signatures in \sn2018gjx}
Calcium emission is observed in \sn2018gjx through the \CaII\ H\&K lines, the [\CaII] \lam\lam 7292, 7324 lines, and the \CaII\ NIR triplet. 
For the duration of our observations, the \CaII\ lines remain typically weaker than the \HeI\ lines.
At earlier times, \CaII\ NIR is the stronger of the calcium features. However, by $+140.7$ days the strongest emission line is [\CaII] \lam\lam 7292, 7324. 
This same evolution is seen in Type IIb SNe \citep[see the relative luminosities in][]{Jerkstrand2015} and \sn2015G, suggesting that the electron density \nel\ in the Ca line-forming regions is relatively similar between these objects.

Strong \OI\ \lam7774 lines are seen in SNe Ibn; these typically have FWHM velocities greater than those of the \HeI\ lines at the same epoch, marking them as forming due to a different process, possibly from the SN ejecta itself.
The S/N is poor in most of our spectra in this region, but a broad bump can be seen around 7700 \AA. This is not as strong as seen is some other SNe Ibn \citep[e.g., OGLE-SN-2012-006;][]{Pastorello2015b}, where it can appear as strong as the \HeI\ lines.

Emission from forbidden oxygen transitions in the earlier phases is not clearly identified, it is only in the final spectrum at $+140.7$ d that the S/N is sufficient to positively identify oxygen. At this phase we can discern \Oneb\ \lam\lam 6300, 6364 with a double peaked emission and an apparent blueshift of 1300 \kms. 
The velocity shift suggests that an unidentified emission feature seen at $\sim 5500$ \AA\ is \Oneb\ \lam5577 at a similar velocity.
Both of these lines are seen in the early nebular phase of SNe IIb \citep{Jerkstrand2015}.
Offset double-peaked \Oneb\ lines are not uncommon in SE-SNe, and there has been much discussion as to their nature \citep[e.g.,][]{Mazzali2005, Maeda2008,Modjaz2008,Taubenberger2009,Mili2010}. 
In many cases, the horned profile matches well the 6300 \AA\ and 6364 \AA\ peaks, but not in every case, suggesting that there may be some element of coincidence.

\begin{figure*}
    \centering
    \includegraphics[scale=0.76]{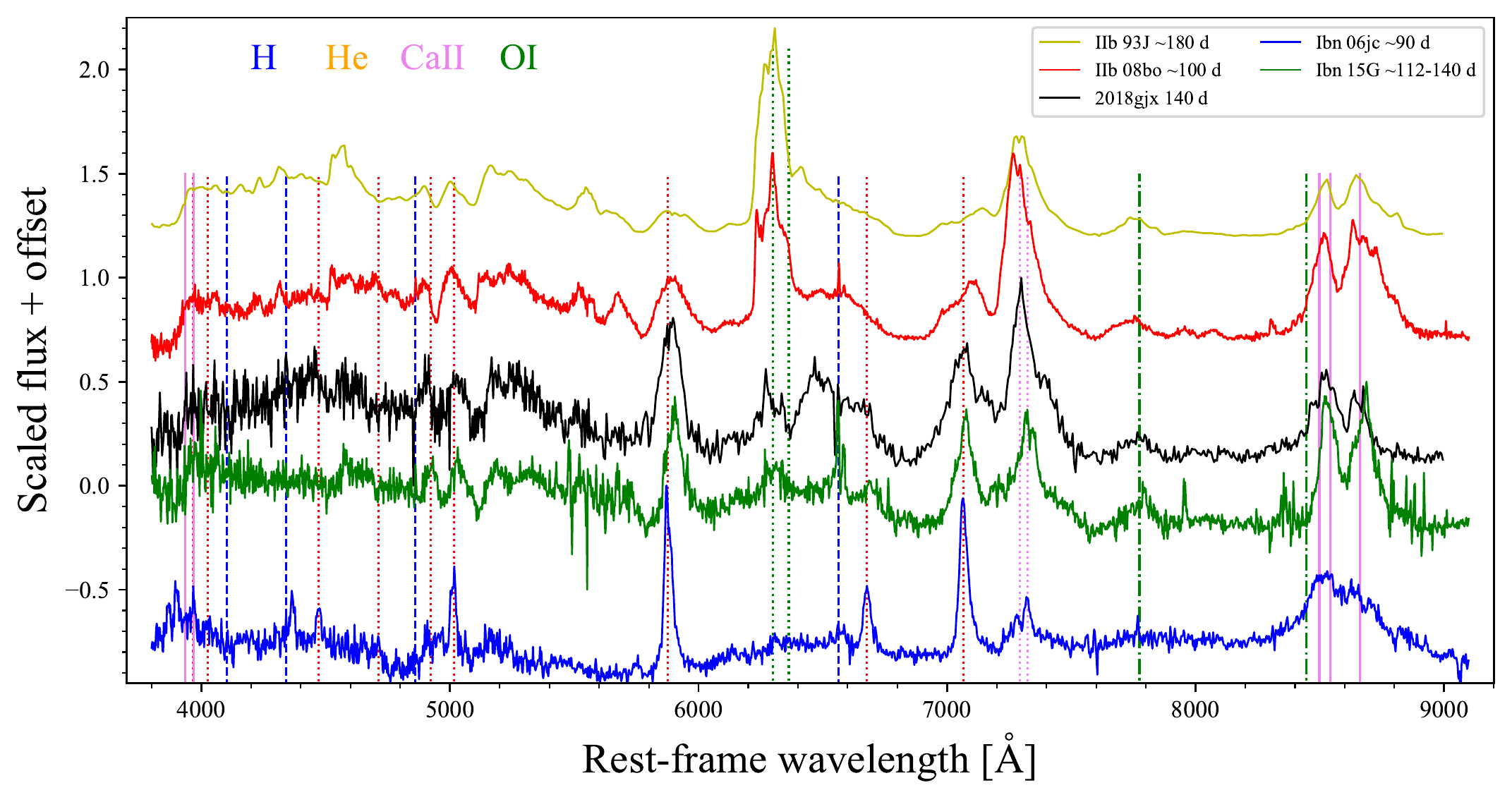}
    \caption{The $+140.7$ d spectrum of SN 2018gjx compared with SNe Ibn 2015G \citep{Shivvers2017b} and 2006jc \citep{Pastorello2008b} and Type IIb \sn1993J \citep{Matheson2000} and \sn2008bo \citep{Modjaz2014}. The spectra of \sn2008gjx, \sn2006jc, and \sn2015G are binned to $\sim 3$ \AA\ for clarity. The key difference between the nebular spectra of SNe IIb and thoseof the interaction dominated SNe Ibn can be seen in the strength of the \Oneb\ \lam\lam 6300,6364 feature, as well as the strength of the \HeI\ lines.}
    \label{fig:Ibncomp}
\end{figure*}

In \sn2018gjx, this line has a luminosity of $<7\times10^{37}$ \ergs, obtained from fitting the region with a double Gaussian to represent the two components. This is considerably weaker than the few $10^{39}$ \ergs\ seen in SNe IIb, including SN\,2008bo, at a similar phase \citep{Prentice2019}.
In H-poor SNe, this line is an effective coolant of the energy deposited from the decay of \Cofs. In most cases it begins to appear a month or so after maximum light and is by far the most prominent line in the spectra by 100 days after explosion (see the comparison in Fig.~\ref{fig:Ibncomp}), with $L$ with a few $\times 10^{39}$ \ergs, considerably stronger than in \sn2018gjx.
We consider why this line appears late and weak in Section~\ref{sec:schematic}.

\section{Modelling}\label{sec:modelling}

\subsection{Modelling of the $+2.4$ d spectrum}\label{sec:specmod}

The first spectrum of SN 2018gjx, obtained 2.4 days after discovery and our only spectrum during Phase I, shows narrow emission lines originating in the CSM. Modelling this spectrum can constrain progenitor properties such as the mass-loss rate, wind velocity, and surface abundances, and also explosion properties at the time of the observation, such as the luminosity and temperature. For this purpose we employ the CMFGEN synthetic spectra from \citet{Boian2020}. CMFGEN \citep{Hillier1998} computes the transport of radiation through spherically symmetric, stationary, expanding atmospheres in non-local thermodynamic equilibrium (non-LTE). For a detailed description of the modelling we refer the reader to \citet{Boian2020}. This library of CMFGEN synthetic spectra covers a wide range of properties, with SN luminosities, $L_{SN}$ from $1.9 \times 10^{8}$ to $2.5 \times 10^{10}$ \lsun, progenitor mass-loss rates, $\dot{M} = 5 \times 10^{-4} - 10^{-2}$ \msunyr (for a terminal wind velocity of $v_{\infty} = 150$ \kms), 3 sets of surface abundances (solar-like, CNO-processed, and He-rich), and SN radii corresponding to 3 epochs ($1$, $1.8$, and $3.7$ d post-explosion).

\begin{figure*}
    \centering
    \includegraphics[scale=0.7]{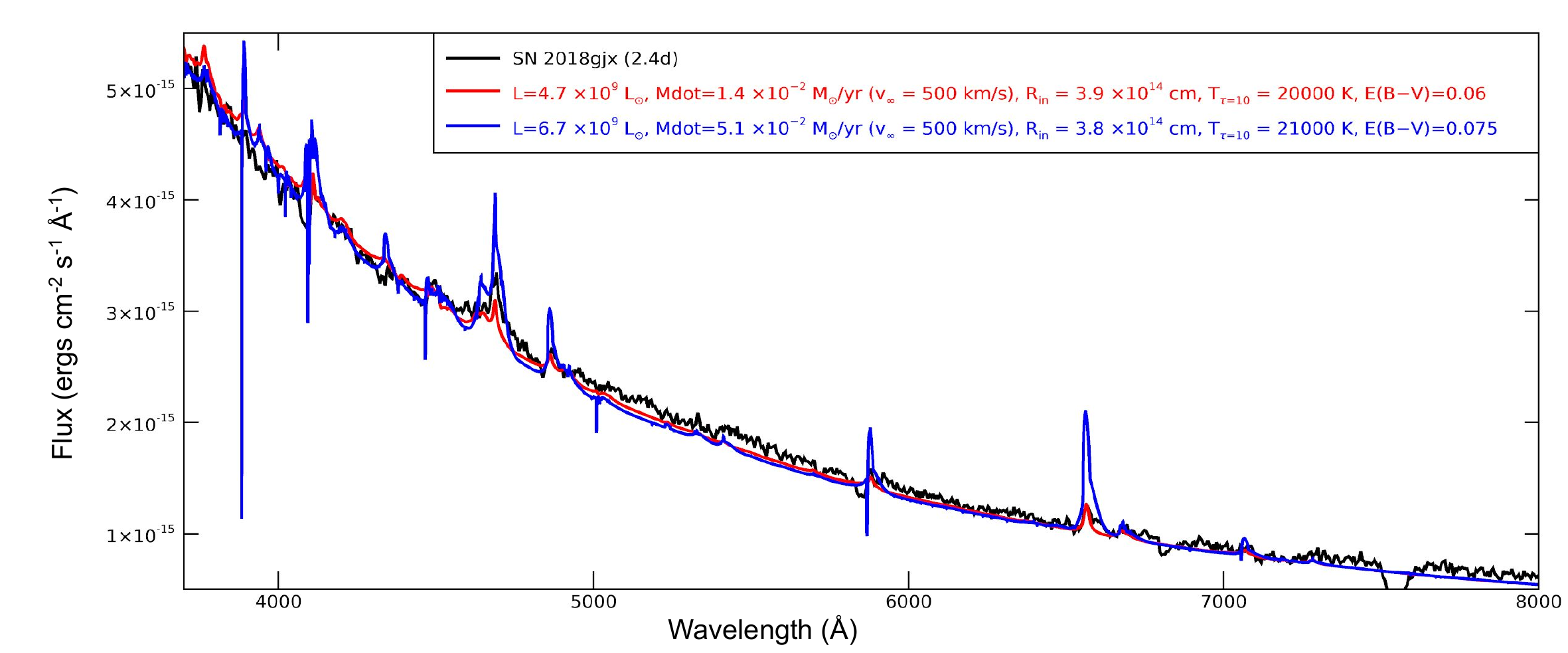}
    \caption{The $+2.4$ day spectrum of SN 2018gjx and the two closest fitting models from \citet{Boian2020} that describe SN ejecta interacting with CSM. The details of the two models are given in the legend and described in more detail in the text. }
    \label{fig:models}
\end{figure*}

The two closest fitting models that encompass the early spectrum of SN 2018gjx have $L_{SN} = 4.7 \times 10^{9} ~\mathrm{L}_{\odot} = 1.8 \times 10^{43}$ \ergs, temperature at an optical depth $\tau=10$ of $T = 20\,000$ K, and $\dot{M} = 1.4 \times 10^{-2}$ \msunyr (red, Fig. \ref{fig:models}), and $L_{SN} = 6.7 \times 10^{9} ~\mathrm{L}_{\odot} = 2.5 \times 10^{43}$ \ergs, $T = 21\,000$ K, and $\dot{M} = 5.1 \times 10^{-2}$ \msunyr (blue, Fig. \ref{fig:models}), respectively. The original terminal wind velocity of the red model is $150$ \kms, but the model spectrum was convolved with a Gaussian with FWHM of $500 $ \kms\ to match the resolution of the observations. The narrow emission lines are not resolved, therefore our best fit values of $v_{\infty}$ and $\dot{M}$ are only upper limits.  The blue model is based on a model from the aforementioned library, but re-computed with $v_{\infty} = 500$ \kms (and a higher $\dot{M}$ to maintain the density of the original model), in an attempt to reproduce the broad absorption component of the \ion{He}{i} $\lambda 5876$ line. Our models cannot reproduce this broad feature, in fact the modelled absorption component originates in inner regions with lower velocity. The broad absorption component may be due to an asymmetric CSM, where high velocity material crosses the line of sight. Both models have He- and N-rich surface abundances. The mass fractions of the elements included in the models, and their values relative to the solar abundances (in brackets), are as follows: H $ = 0.1864$ ($0.26$), He $ = 0.80$ ($2.85$), C $ = 5.58 \times 10^{-5}$ ($1.83 \times 10^{-2}$), N $ = 8.17 \times 10^{-3}$ ($7.42$), O $ = 1.32 \times 10^{-4}$ ($1.38 \times 10^{-2}$). Other elements are also included, with relative masses matching the solar values, totalling $5.2 \times 10^{-3}$. The surface abundances match the expected values for the progenitor of a Type IIb SN. 

The models can also aid in constraining the reddening. Firstly $T$ and $\dot{M}$ are determined using the strength of the lines in the normalised spectrum. Then we compare the observed SED to the best-fit models and redden the synthetic spectra using the \citet{CCM} parametrisation.
The best-fit values for the colour excess, $E(B-V)$ are in the $0.10$ to $0.15$ range, assuming $R_V = 3.1$.  

The best-fit models, assuming a distance to the SN of $35$ Mpc, also reveal an inner radius of $3.8-3.9 \times 10^{14}$ cm ($\sim 5600$ \rsun). If we assume a constant ejecta velocity of $12\, 000 $ \kms, as per the velocities in the IIb-like spectra, this radius corresponds to the spectrum being $\approx 3.7$ days old (w.r.t. the explosion time). This is consistent with the estimated explosion time, considering that the SN ejecta would have initially been slightly higher, and then decelerated by the interaction with the CSM to the value measured from the later spectra.

The interaction is no longer visible in the next available spectrum obtained 23 hours later. This suggests the CSM is confined to a narrow shell-like geometry. If we assume that the spectrum at that stage is $4.36$ days after explosion (upper limit) and take a constant ejecta velocity of $12\, 000$ \kms\ as above, then the CSM extends to at most $4.5 \times 10^{14}$ cm. Further assuming a wind velocity of at most $500$ \kms, the mass-loss occurred over 0.285 years and thus we can place an upper limit on the total mass in the CSM of $\approx 0.4 - 1.4 \times 10^{-2}$ \msun.

\subsection{Estimating $R_0$ from the pseudo-bolometric light curve}\label{sec:lcmod}

In the previous section, the inner boundary of the model was estimated to be $R(3.7 \mathrm{d}) \sim 5600$ \rsun. 
The question arises if we can infer the size of the emitting region at the time of shock-breakout.
To achieve this, the bolometric light curve was fitted using analytical models that account for shock breakout from a stellar surface or wind, as well as a \Nifs-powered component 
in order to constrain the initial photospheric radius $R_0$. 
Based on the late-time spectra that look similar to SNe interacting with He-rich CSM (Type Ibn SNe), we suggest that the bolometric light curve is also likely to have a contribution from interaction with He-rich CSM but this modelling is beyond the scope of our light curve modelling.

\begin{figure}
    \centering
    \includegraphics[scale=0.55]{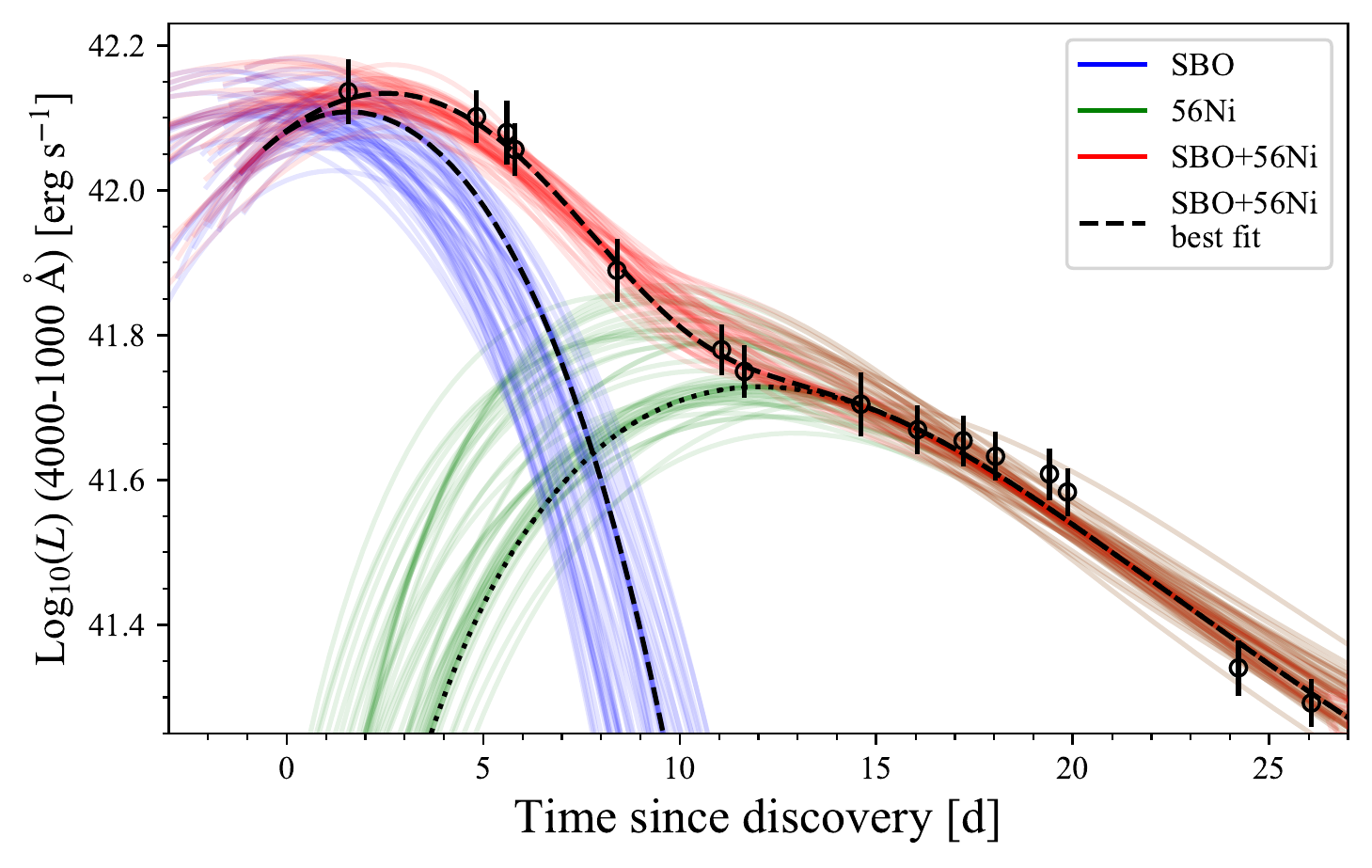}
    \includegraphics[scale=0.55]{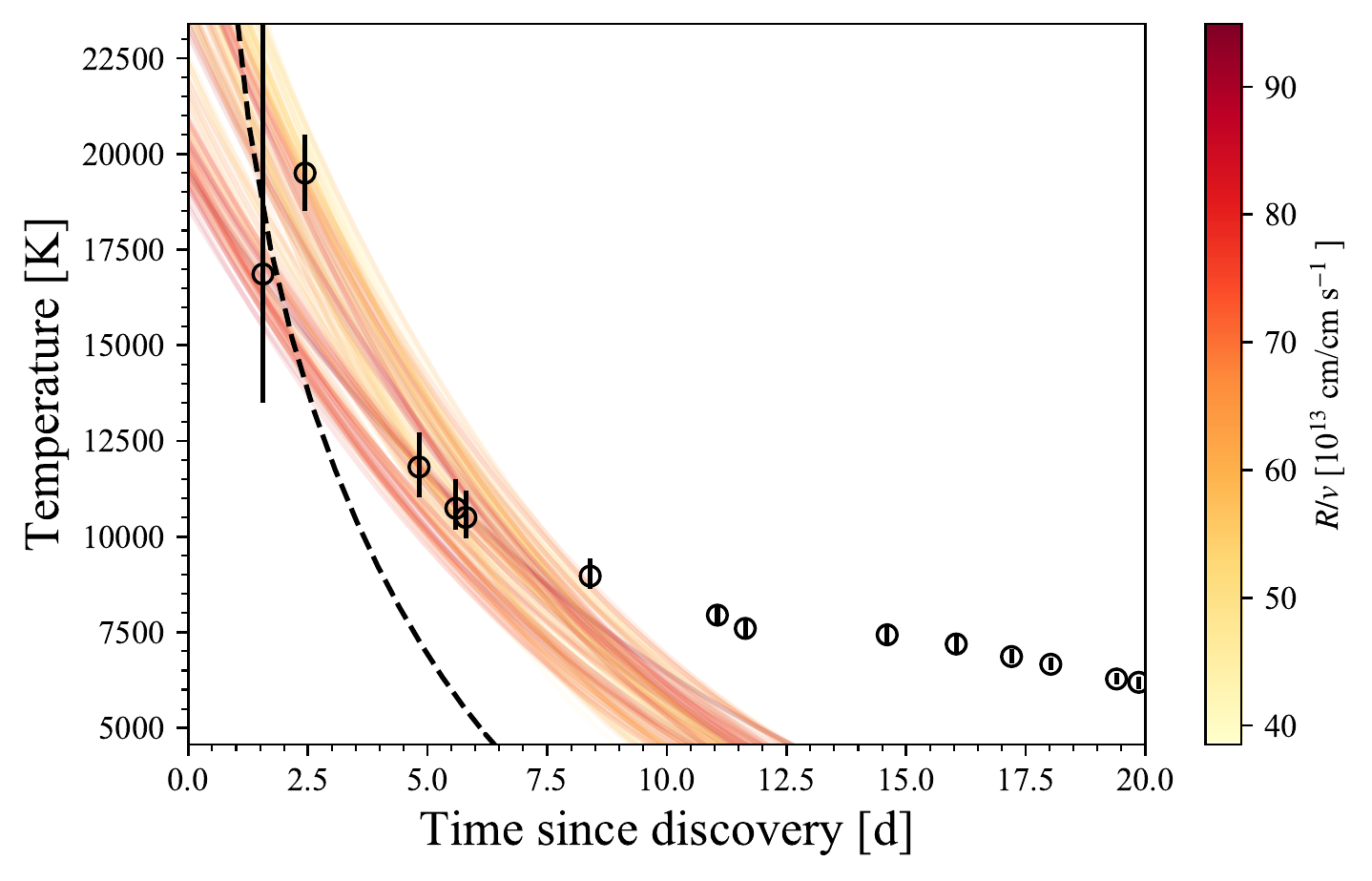}
    \caption{{\bf Top panel:} Fits to the pseudo-bolometric light curve using different models at different phases. Phase I consists of a shock-breakout model (blue), while phase II is a \Nifs\ powered model (green). The addition of the latter helps constrain the properties of the former. The sum of the two is shown in red. {\bf Lower panel:} The temperature evolution for the early phases as given by the shock-breakout model plotted against the observed $T$, derived from SEDs and $+2.4$ d spectrum, assuming blackbody emission. The temperature of this blackbody constrains the ratio of radius/velocity that produce fits to the light curve. Here, the colourmap corresponds linearly to 40 -- 95 [$10^{13}$ cm]/[cm s$^{-1}$]. For smaller ratios, the temperature decay rate is too rapid to match the observed temperature, as demonstrated by the black dashed line for fixed $v_\mathrm{e}=14000$ \kms. After 7.5 days, the temperature becomes increasingly dominated by energy deposited from the \Nifs\ decay chain and the validity of the temperature estimate breaks down }
    \label{fig:LCfits}
\end{figure}

To model the early light curve of \sn2018gjx in Phase I, we have used the analytical shock-breakout model of \citet{Piro2015}, as expressed in \citet{Arcavi2017} for the Type IIb \sn2016gkg \citep{Tartaglia2017}.
Contribution of a \Nifs\ powered component \citep{Arnett1982,Valenti2008} is included in order to constrain the early emission.
In fitting the data, the explosion date $t_0$ was allowed to vary between 0.5 and 4.5 days before detection. This latter limit is earlier than the ATLAS non-detection but is necessary in order to allow for the possibility of a small initial radius but high initial temperature in the early evolution of the light curve. This emission would be UV bright, but dim in $o$.  

The SBO is modelled in terms of core mass \Mc, the envelope mass \Me, the envelope (reverse shock) velocity \ve\footnote{Note that \ve\ is the reverse shock velocity and defines the envelope expansion rate, it is not equivalent to the forward shock velocity $v_s$.}, and the initial radius of the stellar envelope $R_0$. 
The model is quite insensitive to \Mc, and degenerate between \ve, $R_0$, and \Me. 
In this model the luminosity $L(t)$ is assumed to result from a black-body continuum of $T(t)$. To fit the pseudo bolometric light curve we take a Planck function with $T(t)$ and scale it so that its integrated luminosity over 1--1\,000\,000 \AA\ is equal to $L(t)$. 
The scaled Planck function is then integrated over a reduced range of 4000--10000 \AA\ to obtain the luminosity over the same range as the pseudo-bolometric light curve.
The final check compares the temperature with the observed temperature evolution of the transient. This breaks degeneracies found in some of the parameters, and rules out high shock velocities.

From the fitted light curve (shown in Fig.~\ref{fig:LCfits}), we obtain a low velocity fit where \ve\  $ = (0.30\pm{0.04})\times 10^{9} $ cm s$^{-1}$ ($\sim 3000$ \kms), with an extended  $R_0=(21\pm{3})\times 10^{13}$ cm ($\sim3000$ \rsun), but low mass \Me\ $=0.06\pm{0.01}$ \msun\ envelope. 
The fits give a time of explosion $t_0 = 3\pm{1}$ day. 
After 2--3 days, the extent of the photospheric radius \rph\ is comparable to the inner boundary of the spectroscopic model from the previous section. 
Whilst a ``high velocity'' fit is possible, as is shown in the lower panel of Fig.~\ref{fig:LCfits}, the temperature evolution is inconsistent with small $R_0$ and large \ve\, which requires a large initial temperature and a steep temperature gradient to match the light curve. 

The importance of this result is that the estimated photospheric radius is several times greater than even the largest known stars, which supports the argument that the early emission is not from a stellar surface, but rather from extended material around the star.

\subsubsection{Properties of the \Nifs\ model}
The \Nifs\ model is primarily used to constrain the SBO model, however we can estimate some physical parameters from it.
The fit is constrained to have the same $t_0$ as the SBO model and rises to peak in $\sim 13$ days with \mni\ $=0.021\pm{0.001}$ \msun.
Using the simple \mej\ estimate of \citet{Arnett1982}, a rise time of 13 days and characteristic velocity \vsc\ of 5000--8000 \kms\ from the measured \FeII\ and \HeI\ velocities at this phase we find an ejecta mass of $\sim1.0-1.5$ \msun.
Again, the assumption here is that the contribution from interaction to powering the light curve is significantly less than that of the \Nifs\ decay chain.
These values would place \sn2018jx at the lower end of the parameter distributions for SNe IIb given in \citet{Prentice2019}, and similar to \sn2008bo.

\section{Discussion}\label{sec:discussion}
\sn2018gjx is an unusual event, which extends the diversity of not only SNe interacting with He-rich CSM but also stripped envelope SNe.
Any interpretation of the transient must be self consistent. It must be able to explain:

\begin{enumerate}
    \item {The initial hot and blue spectrum with \HeII\ emission lines and P-Cygni features, as well as the observed velocities, and the estimated photospheric radius.}
    \item{The SN IIb phase, where the spectra are dominated by broad SN features consistent with SNe IIb, as well as the velocities that are measured during this phase}
    \item{The evolution to an emission-dominated phase after one month that is similar to those seen in SNe Ibn, where there is interaction with He-rich CSM and the features expected in the nebular phase for SE-SNe and in particular SN IIb, are not seen (e.g., there is no strong \Oneb\ emission)}

\end{enumerate}

\subsection{An interpretation of the progenitor system}\label{sec:schematic}

\begin{figure*}
    \centering
    \includegraphics[scale=0.25]{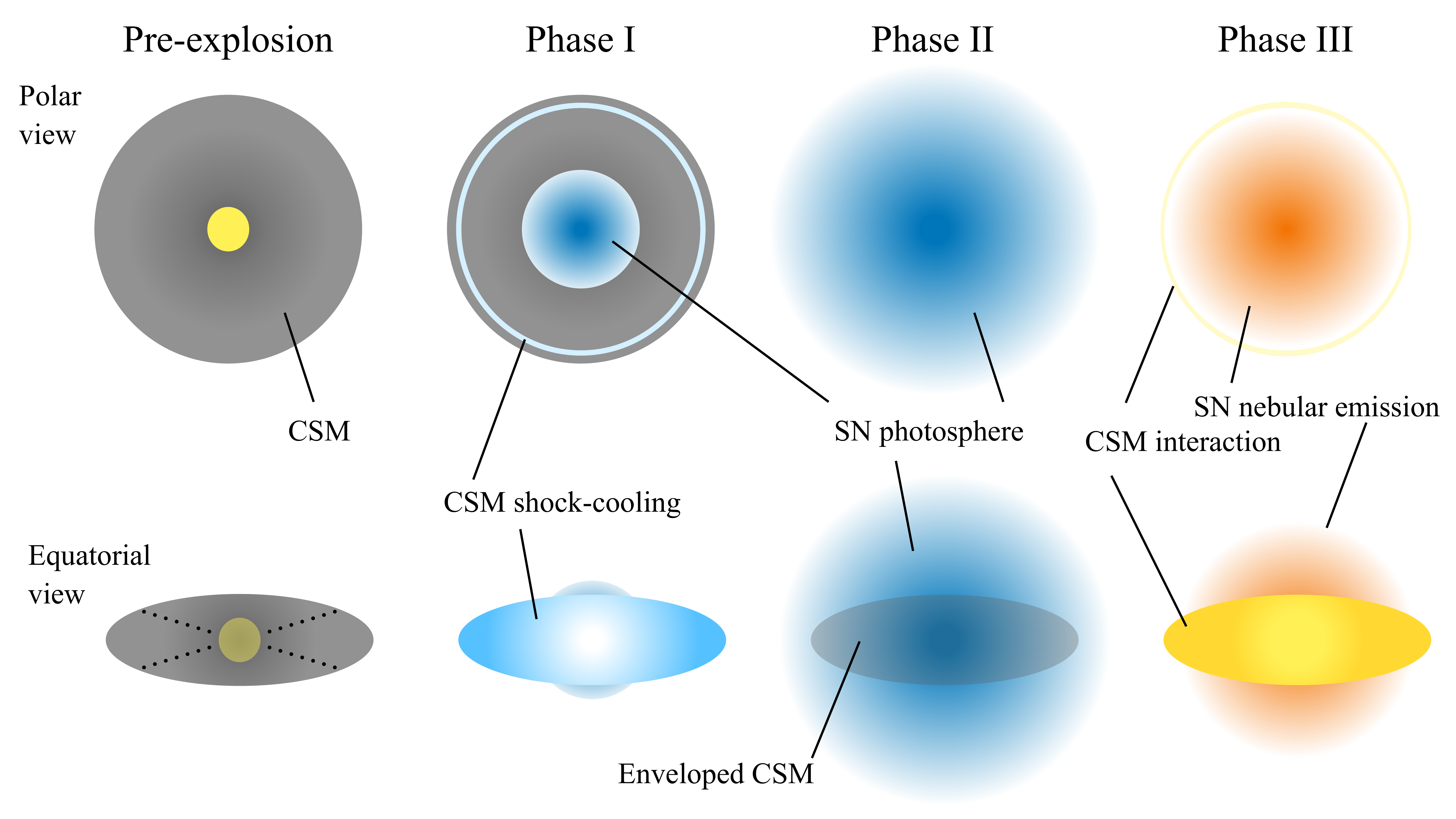}
    \caption{Cartoon schematic, not to scale, of a proposed system configuration to explain the properties of \sn2018gjx, based upon a model suggested for SNe IIn by \citet{Smith2015b}. Polar and equatorial views are shown. The grey disk represents a He-rich CSM torus, the densest part of an aspherical distribution of CSM for which the rest is not shown. The dotted lines in the equatorial view show a vertical slice through the system to demonstrate that the bulk of the CSM is equatorial, leaving the poles less obscured by material. In Phase I the star explodes. The polar observer sees mostly SN and little shock-breakout from the CSM, the equatorial observer sees mostly shock-breakout. In phase II the SN photosphere envelops the CSM leaving both observers to see a SN\,IIb. As the photosphere recedes in velocity space, the equatorial observer sees more contribution from CSM interaction. The presence of the CSM presents a region of higher opacity to photons from within the ejecta, possibly resulting in a dimming effect equatorially. In Phase III the SN transitions to the nebular phase and the polar observer sees the entire nebular emission from the SN with a strong \Oneb\ \lam\lam6300, 6364 line. The equatorial observer however primarily sees the interaction region, the observed \Oneb\ line is weak because these photons scatter off the intervening CSM and escape elsewhere. }
    \label{fig:schematic}
\end{figure*}

Here we propose a possible solution which may be able to explain the key observables, which is illustrated in Fig.~\ref{fig:schematic}. We consider a compact and/or stripped progenitor star, which has recently undergone a mass-loss event, either through outburst or a wind, so it is surrounded by a dense optically thick CSM. 
As observed with massive stars in our own Galaxy, the distribution of this material is not spherical, but has an aspherical distribution around the progenitor, which may be a disk, torus, or polar outflow \citep[e.g.,][]{Smith2002,Grundstrom2007,Vink2007,Smith2013,Smith2015b,Smith2017}. 

We consider a situation where the distribution of CSM is aspherical, with the bulk located in a torus, as proposed for the Type IIn, PTF11iqb, by \citet{Smith2015b}, and that our view of the system is such that the torus is between us and the star.
When the star explodes, our view of the actual stellar surface is obscured. 
The outermost SN ejecta sweeps through the CSM, which leads to weak interaction at the ejecta/CSM interface. This, combined with the deposition of shock energy into the material, leads to high temperatures that ionises some of the surrounding unshocked CSM \citep{Boian2020}.
The winds from hot, compact, massive stars can reach over 1000 \kms, in agreement with the velocities measured from the narrow P-Cygni features present in the first spectrum ($+$2.4 d from explosion), this is Phase I.

The observed low luminosity, blue to red colour evolution, and lack of strong narrow emission lines in the spectra is enough to suggest that strong interaction is not occurring at this time (as compared with \sn2006jc).
This scenario is sufficient to explain the spectroscopic and photometric appearance of the earliest phases as detailed item (i) in the introduction to Section \ref{sec:discussion}.

In Phase II (10 -- 30 d), the ejecta photosphere (\rph$=50-80\times10^{13}$ cm for $T= 6000-7500$ K and $L=8\times10^{41}$ \ergs) envelops the material in the torus, which explains the outward spectroscopic appearance of a SN IIb. 
Precedence for the luminosity of this event has already been set by the low luminosity \sn2008bo, however, if one assumes that the majority of \Nifs\ synthesised in the explosion is roughly spherically symmetric and centrally located, then the thermalised photons from this region may preferentially escape from the polar regions where the material density is lower, and only some fraction will contribute to heating the torus.
The case of \sn2008bo means we do not require significant luminosity-boosting for a polar observer, as would be the situation if all SNe IIb were 93J-like.
This scenario can explain item (ii), the SN IIb phase, where the spectra are dominated by broad SN components and the pseudo-bolometric light curve is powered predominantly by \Nifs\ decay.

As time passes, the SN photosphere recedes in velocity space until the interaction region begins to reappear, this is already demonstrated by the growing strength of the \HeI\ lines in the spectra through Phase II.
Depending on the thickness of the torus and its rate of expansion, if it remains optically thick due to ongoing interaction it could present a sufficiently large solid angle so as to obscure the lower velocity inner regions of the SN ejecta.
During the epoch covered by Phase III of {$>+30$ d}, a SN would typically enter the nebular phase.
We assume this is the case here, but the inner low-velocity O-rich material that produces the \Oneb\ \lam\lam 6300, 6364 line could be mostly obscured to an equatorial observer. 
We can justify this by considering that the FWHM of the \Oneb\ line in SE-SNe is typically around 5000 -- 6000 \kms\ for the combined 6300 \AA\ and 6364 \AA\ components of multiple distributions of \Oneb\ \citep{Taubenberger2009}, which is comparable to the FWHM of the \HeI\ lines measured for \sn2018gjx. This means that the majority of the \Oneb\ emitting material can be found within the region bounded by the CSM.
Photons from this region can escape through the optically thin ejecta at the poles however. 
Photons that scatter into the torus region can contribute to heating, which explains why the tail follows that of \Cofs\ decay, but they can equally be scattered back into the optically thin region and escape. This serves to mask the \Oneb\ emission line and prevents the flux from this line contributing significantly the visible spectrum.
The radiated energy at this time comes from some contribution of interaction and \Cofs\ decay.
This explains item (iii) on the list.

We make no predictions as to the thickness of the torus or the inclination of the system, other than to say that the interaction signatures require it to be aligned somewhat in our direction. We only require that such a system is in a configuration for us to see the transient in the way it is.

\subsection{SN\,2018gjx in the context of SN types}
\sn2018gjx presents a rare opportunity to view the underlying explosion of a Type Ibn.
Conventionally, interacting SNe hide the normal SN light through a mix of high luminosity and opacity.
In this case, the SN ejecta away from the interaction region overtakes the location of the CSM, providing us with an opportunity to see the ejecta photosphere and showing us how it relates to other known object.

\subsubsection{Type IIb}
The underlying SN properties of \sn2018gjx, aside from the low luminosity, are normal for Type IIb SNe.
The estimated metallicity at the explosion site is also typical for this SN type \citep{Kuncarayakti2018}.
If our hypothetical model of the system configuration is correct, then it may be that many observed SNe IIb are events similar to this with a different viewing angle and varying CSM densities.
SNe IIb are known to interact with CSM eject at late times and the radio light curves suggest that many do without the interaction being visible in the optical \citep{Chevalier2010}.
\sn2015G tantalisingly suggests that that some SNe Ib may also find themselves in such a configuration.
We suggest that this sub-population, where interaction dominates at late times only, may have been missed due a mixture of chance, the inherent low luminosities of SNe IIb \citep{Prentice2019}, and the lack of objects followed past maximum light. 
We anticipate that there may be unpublished data in existence of 18gjx-like objects, but which have not been identified for their unusual nature.

\subsubsection{Type Ibn}
SNe Ibn display heterogeneity across their photometric and spectroscopic evolution, they vary significantly in rise time, late decay time, and peak luminosity.
\citet{Hoss2017} analysed the light curves and spectroscopic evolution of around 25 SNe Ibn. They found that the light curves were broadly homogeneous for the first month after maximum light with a decline rate of $0.05- 0.15$ mag day$^{-1}$.
\sn2018jx peaks in the optical some three days after explosion, and then shows a variety of decline rates owing to its shock-cooling phase. If we instead consider the period around 15-20 days after explosion as peak then we find that for a brief period 20--30 days after explosion, \sn2018gjx matches this decline rate (Fig.~\ref{fig:LCs}).
\citet{Hoss2017} find that SE-SNe decay slower on average, although their light curves also tend to be broader. However, \sn2008bo, very clearly a Type IIb, displays a similar light curve to that of \sn2018gjx.
Finally, with a peak $r$ band magnitude of approximately $-17.2$ mag, \sn2018gjx is at the lowest part of their luminosity distribution.

In terms of spectroscopy, the results of \citet{Hoss2017} show that there is spectroscopic diversity at maximum light and they suggested that SNe Ibn could be divided into two groups; those that displayed P-Cygni profiles (e.g., SN 2010al, 2015G) and those that showed emission (e.g., SN 2006jc).
\sn2018gjx would be considered part of the ``P Cygni'' class based upon our first spectrum but this is a loose classification, because the transient then passes through a Type IIb phase before showing spectroscopic signatures reminiscent of Type Ibn.

\subsubsection{Assessing the progenitor star}
The progenitors of SNe IIb have been inferred to be, from pre/post-explosion imaging, yellow supergiant (YSG) or red supergiant (RSG) stars \citep[see, for example,][]{Aldering1994,Maund2011,VanDyk2014,Groh2014}.
This has contrasted with the results of radio observations \citep{Soderberg2012}, and stellar evolution models \citep{Groh2013b} which suggest more compact, Wolf-Rayet stars for some of these events.
Binary evolution is often presumed in order to reconcile the low ejecta masses for these events with stellar evolution models.
The progenitor of \sn2018gjx may well be stripped through binary interaction, which in turn would favour the CSM distribution used in the hypothesised scenario. 
Our results from modelling of the Phase I spectrum suggests a compact progenitor with a mixed H/He layer and an extended CSM around it; a Wolf-Rayet or stripped LBV star based upon the models of \citet{Groh2013b}. 
The measured line velocities in this spectrum also disfavours a Y/RSG progenitor.
The inferred properties of the progenitor from the models of the early spectra, with a 20\% surface abundance of H, suggests that this is a relatively rare type of phase in which for a star to explode.
\sn2018gjx is an unusual event compared to known types of stellar explosions, so its progenitor properties may not be representative of either Type IIb or Type Ibn. It does, however, provide further evidence towards a compact progenitor. 
The question following then arises -- given that CSM greatly affects the observables of the explosion, could it also affect the pre-explosion colour of an apparent progenitor, even if the CSM is not directly aligned along the line of sight, with the effect of placing the progenitor star in an different part of the HR diagram?

\section{Conclusions}
\label{sec:conclusions}
We have presented the photometric and spectroscopic evolution of the unusual transient \sn2018gjx, which passes through three distinct phases.
Phase I shows a rapidly declining light curve and a hot, blue spectrum showing weak P-Cygni features and UV-ionised \HeII\ lines. 
The light curve levels off during Phase II ($\sim+10$ to $+30$ d) and spectra are similar to SNe IIb at maximum light.
Finally, in Phase III the spectra evolve into the characteristic spectra of a subset of SNe Ibn which display a photospheric phase, with strong and relatively broad He and Ca lines, but an absence of \Oneb\ \lam\lam 6300, 6364 emission, as seen in SNe IIb.
The \HeI\ lines are estimated to have a FWHM of $v \sim 5500$ \kms\ that persists until at least $+140$ days.

The measured line velocities during Phase II are consistent with SNe IIb at around 7000 -- 12\,000 \kms. 
In Phase III we find that the FWHM of the \HeI\ emission lines remain almost constant at $\sim5500$ \kms, which is counter to the velocity evolution seen in the \HeI\ lines of the few SNe Ibn observed as late as \sn2018gjx.
An unusual broad feature seen around 6560 \AA\ in the Phase III spectra could be N emission from the He shell, or it could be evidence of interaction of H in a shell-like structure with CSM.
Our last spectrum reveals the \Oneb\ \lam\lam6300, 6364 emission line, but its luminosity is 50--100 times lower than that seen in SNe IIb. 

Models of the Phase I spectrum at $+2.4$ d gives an estimated mass loss of $\dot{M} = (1.4-5.1) \times 10^{-2} M_{\odot}$ yr$^{-1}$
and CSM mass of $(0.4 - 1.4) \times 10^{-2} ~M_{\odot}$. The models find a Wolf-Rayet or stripped LBV like He and N rich surface abundance with 20\% mixed H, which supports the supposition that argument for the origin of the broad 6560 \AA\ feature.
From the spectroscopic modelling of the $+2.4$ d spectrum an inner boundary to the emitting radius is found to be $\sim 5600$ \rsun, and an initial radius of $R_0 >3000$ \rsun\ from applying an SBO model to the pseudo-bolometric light curve.
This radius is considerably larger than stellar radii, and is further evidence of significant CSM around a more compact progenitor.

We suggest a model, adapted from one proposed for SNe IIn, whereby the progenitor star explodes within an aspherical CSM, probably a torus. Our viewing angle is such that the explosion itself is obscured by the CSM, but which itself is not dense enough to result in the luminosities seen in typical interacting SNe.
The SN photosphere overtakes the interaction region briefly, giving rise to the appearance of a SN IIb.
Once the photosphere recedes, the interaction region becomes the dominant spectroscopic feature and the inner workings of the SN (i.e., transition to the nebular phase) are hidden from us, resulting in weak \Oneb\ lines. 
The relatively low luminosity of the transient is explained through the heterogeneous density profile of the system, with the poles having a lower density than the equatorial region allowing photons to more readily escape there.
This configuration is viewing angle dependent. A polar observer may see a normal SN IIb, while an observer with the CSM directly in their line of sight may never see the actual SN and instead may see a SN Ibn. Given that the proposed scenario is highly aspherical, future observations of 18gjx-like events should include polarimetry in order to test the degree of asphericity in the emission.

18gjx-like events would be difficult to identify based upon a single spectroscopic observation, which demonstrates that snapshot classification misses diversity within the transient population. This should be considered when planning how to overcome the ``classification gap''; the difference between the number of discoveries and the number of classifications.

\section*{Acknowledgements}
SJP and KM are supported by H2020 ERC grant no.~758638.
Research by K.A.B, S.V., and Y.D. is supported by NSF grant AST-1813176.
DAH, DH, JB, and CP are supported by NSF grants AST-1911225 and AST-1911151, and NASA grant 1518168. 
C.M. was supported by NSF AST-1313484.
This paper made use of data from the LCO network.
L.G. was funded by the European Union's Horizon 2020 research and innovation programme under the Marie Sk\l{}odowska-Curie grant agreement No. 839090, and by the Spanish grant PGC2018-095317-B-C21.
CPG acknowledges support from EU/FP7-ERC grant no. [615929].
MG is supported by the Polish NCN MAESTRO grant 2014/14/A/ST9/00121.
TMB was funded by the CONICYT PFCHA / DOCTORADOBECAS CHILE/2017-72180113.
MN is supported by a Royal Astronomical Society Research Fellowship.
SJS acknowledges funding from STFC Grant Ref: ST/P000312/1.
H.K. was funded by the Academy of Finland projects 324504 and 328898.
L.T. acknowledges support from MIUR (PRIN 2017 grant 20179ZF5KS).
The Liverpool Telescope is operated on the island of La Palma by Liverpool John Moores University in the Spanish Observatorio del Roque de los Muchachos of the Instituto de Astrofisica de Canarias with financial support from the UK Science and Technology Facilities Council.
Based on observations collected at the European Organisation for  Astronomical Research in the Southern Hemisphere under ESO programme 199.D-0143.
This work has made use of data from the Asteroid Terrestrial-impact
Last Alert System (ATLAS) project. ATLAS is primarily funded to search
for near earth asteroids through NASA grants NN12AR55G, 80NSSC18K0284,
and 80NSSC18K1575; byproducts of the NEO search include images and
catalogs from the survey area.  The ATLAS science products have been
made possible through the contributions of the University of Hawaii
Institute for Astronomy, the Queen's University Belfast, and the Space
Telescope Science Institute. 
Some of the data presented herein were obtained at the W. M. Keck Observatory, which is operated as a scientific partnership among the California Institute of Technology, the University of California, and NASA; the observatory was made possible by the generous financial support of the W. M. Keck Foundation.
Partly based on observations made with the Nordic Optical Telescope, operated at the Observatorio del Roque de los Muchachos, La Palma, Spain, of the Instituto de Astrofisica de Canarias.
ALFOSC is provided by the Instituto de Astrofisica de Andalucia (IAA).
We thank the anonymous referee for their time and comments, which improved this manuscript.

\section*{Data Availability}
Data will be made available on the Weizmann Interactive Supernova Data Repository (WISeREP) at https://wiserep.weizmann.ac.il/.




\bibliographystyle{mnras}
\bibliography{18gjxbib} 




\appendix


\bsp	
\label{lastpage}
\end{document}